# Liquidity Premium, Liquidity-Adjusted Return and Volatility, and Extreme Liquidity

Qi Deng[1,2,3,4*]
Zhong-guo Zhou[5]

## Abstract

We establish innovative liquidity premium measures, and construct liquidity-adjusted return and volatility to model assets with extreme liquidity, represented by a portfolio of selected crypto assets, and upon which we develop a set of liquidity-adjusted ARMA-GARCH/EGARCH models. We demonstrate that these models produce superior predictability at extreme liquidity to their traditional counterparts. We provide empirical support by comparing the performances of a series of Mean Variance portfolios.

Funding Source: The work was supported by Hubei University of Automotive Technology [grant number BK202209] and Hubei Provincial Bureau of Science and Technology [grant number 2023EHA018].

1. College of Artificial Intelligence, Hubei University of Automotive Technology, Shiyan, China
2. Jack Welch College of Business & Technology, Sacred Heart University, Fairfield, CT, USA
3. School of Accounting, Economics and Finance, University of Portsmouth, Portsmouth, UK
4. Cofintelligence Financial Technology Ltd., Hong Kong and Shanghai, China
5. Department of Finance, Financial Planning, and Insurance, Nazarian College of Business and Economics, California State University Northridge, CA, USA
*. Corresponding author: dq@huat.edu.cn; dengq@sacredheart.edu

**Liquidity Premium, Liquidity-Adjusted Return and Volatility, and Extreme Liquidity**

**1. Introduction**

The existing literature, in general, is not adequate in modeling assets that exhibit exhibit extreme liquidity. The recent enormous rise of crypto assets aggravates such inadequacy. Thus, the goal of this paper is to develop a theoretical and methodological framework with embedded treatments of liquidity risk, which is particularly effective in modeling assets with extreme liquidity. There are three reasons why we choose crypto assets to demonstrate our framework. First and at the market level, as the most visible representation of unregulated assets with extreme liquidity, the most dominant cryptos (i.e., BTC and ETH) are becoming more efficient (Urquhart, 2016; Nadarajah and Chu, 2017; Tiwari et al., 2018) from an Efficient Market Hypothesis (EMH) viewpoint (weak form). Second and at the asset level, more investors are now willing to include crypto assets in their portfolios (Koutmos, King and Zopounidis, 2021). Third and from a technicality perspective, trading of crypto assets is not (yet) subject to regulatory trading restrictions (no return caps, nor price or volume triggered trading curbs), as such, the effect of (extreme) liquidity is more prominent in return predictability (Wei, 2018).

The most widely used measure of asset liquidity is the "Illiquidity Ratio (ILR)" proposed by Amihud (2002).[1] In this paper, we use the tick-level trading data of ten selected crypto assets to establish a scaled minute-level illiquidity ratio, derive the minute-level liquidity-adjusted return and volatility, and introduce a minute-level liquidity measure, the liquidity premium Beta. We further calculate their day-level equivalents that include a daily liquidity premium Beta, which reflects daily liquidity. We find that the distribution of daily liquidity of crypto assets is positively skewed with a wide and sticky distribution, which brings in new information from a third dimension of liquidity, in addition to the existing dimensions of return and variance. The highly asymmetric distribution of liquidity transfers the distribution of daily regular return and (intraday) volatility into a wider distribution of daily liquidity-adjusted return and (intraday) volatility, which is "tighter" when liquidity is high, but "looser" when liquidity is low.

---

[1] https://www.stern.nyu.edu/experience-stern/news-events/professor-yakov-amihud-s-illiquidity-ratio-referenced

It is widely accepted that the traditional ARMA-GARCH/EGARCH structures are in general not effective in modeling assets with price jumps (e.g., Liu, Liu and Zhang, 2019) as their return and volatility no longer exhibit well-behaved autoregressive properties because of the jumps. This prompts researchers to explore other statistical models that use realized volatility measures, such as the HAR (Corsi, 2009). We hypothesize that, asset price jumps are manifestations of extreme liquidity, thus if we adjust the return and volatility (of assets with jumps) with liquidity measures, their well-behaved autoregressive properties would be restored, and the AMRA-GARCH/EGARCH would become effective again in modeling these assets. We model the daily conditional liquidity-adjusted return and (intraday) variance with the ARMA-GARCH/EGARCH, which, on theory, produce reduced forecasting error under extreme liquidity, therefore provide much improved predictability.[2] To provide empirical evidence for such improved predictability, we compare the performances of a series of LAMV and TMV portfolios. We observe clear advantage for the LAMV over the TMV in portfolio optimization,[3] with or without the treatments on potential wash trades that count for a substantial portion in crypto trading (Cong et al., 2023).

---

[2] We use the standard ARMA-GARCH/EGARCH to model the conditional liquidity-adjusted return and (intraday) volatility, based on which we develop a set of new ARMA-GARCH/EGARCH models with built-in liquidity terms, which we name the "liquidity-adjusted ARMA-GARCH/EGARCH" models. We also use the standard ARMA-GARCH/EGARCH to model the conditional regular (non-liquidity-adjusted) return and (intraday) volatility, and we call these models "traditional ARMA-GARCH/EGARCH" models. We use these terms throughout this paper.

The contribution of this paper is two-fold. First, we derive and formulate liquidity-adjusted return and volatility, and liquidity measures at both minute and daily levels. The daily liquidity is highly asymmetric with a wide, positively skewed, and sticky distribution, resulting in a highly asymmetric ensemble distribution of liquidity-adjusted return and volatility. Second, the highly asymmetric distribution of liquidity-adjusted return and volatility enhances the liquidity-adjusted ARMA-GARCH/EGARCH models with much improved predictability at extreme liquidity, and we develop a set of LAMV portfolios to empirically demonstrate such improved predictability with selected crypto assets. While we use crypto assets to demonstrate our models, in general, they are effective on other assets with extreme liquidity.

The rest of the paper proceeds as follows. Section 2 reviews existing literature on liquidity, volatility of liquidity, and models of assets with extreme liquidity. Section 3 introduces the liquidity measures and liquidity-adjusted return and volatility, at both minute and daily levels. Section 4 provides the descriptive statistics of the data and discussions on the ensemble distribution of liquidity-adjusted return and volatility. Section 5 presents the improved predictability of the liquidity-adjusted ARMA-GARCH/EGARCH models. Section 6 constructs the LAMV portfolios with selected crypto assets to provide empirical evidence for the improved predictability. Section

---

[3] We use the standard Mean-Variance (MV) method to optimize portfolios with forecasted returns produced by the liquidity-adjusted ARMA-GARCH/EGARCH models, and name these portfolios as "liquidity-adjusted Mean Variance (LAMV)" portfolios. We also use the standard MV to optimize portfolios with forecasted returns produced by the traditional ARMA-GARCH/EGARCH models, and name these portfolios as "traditional Mean Variance (TMV)" portfolios. We use these terms throughout this paper.

7 concludes the paper. We also include an Appendix that gives the details of deriving the univariate liquidity-adjusted ARMA-GARCH/EGARCH models.

## 2. Literature Review on Asset Liquidity

Amihud and Mendelson (1986) define an asset's illiquidity as the cost of immediate execution (trading friction) and use bid-ask spread as a general measure of illiquidity. However, some large traders may trade large blocks of shares outside the bid-ask spread (Chan and Lakonishok, 1995; Keim and Madhavan, 1996), and while the bid-ask spread explains the returns of NASDAQ, it does not explain that of NYSE and AMEX (Eleswarapu, 1997), thus some literature uses price impact (return) as a liquidity measure (for example., Brennan and Subrahmanyam, 1996; Amihud, 2002; Pástor and Stambaugh, 2003; Acharya and Pedersen, 2005). Amihud, Mendelson and Pedersen (2005) survey academic literature on illiquidity and its linkage to required return. As their survey is extensive and is in particular relevant to our study, we follow its classification of liquidity pricing models with our own interpretations in this literature review.

### 2.1 Illiquidity Proxies

Fong, Holden and Trzcinka (2017) identify a multitude of high-quality liquidity proxies and find that the daily version of Closing Percent Quoted Spread (Chung and Zhang, 2014) is the best daily percent-cost proxy, and that the daily version of Amihud (2002) is the best daily cost-per-dollar-volume proxy, which is given as:

$$ILR_t = \frac{1}{D_t} \sum_{\tau=1}^{D_t} \frac{|r_\tau|}{V_\tau} \qquad (1a)$$

$$ilq_\tau = \frac{|r_\tau|}{V_\tau} \qquad (1b)$$

where $ILR_t$ is the average illiquidity ratio in year t, $D_t$ is the number of trading days in a year,
$ilq_\tau$ is the illiquidity ratio on day $\tau$, $r_\tau$ is the return on day $\tau$, $V_\tau$ is the dollar-adjusted volume traded on day $\tau$.

As we use minute-interval trading data of selected crypto assets, the daily version of Amihud (2002) given in Equation 1b is a natural starting point for building our own illiquidity measures, which are essentially a cost-per-dollar-volume variation.

**2.2 Liquidity Costs and Pricing Models**

Acharya and Pedersen (2005) present a simple equilibrium model, the liquidity-adjusted capital asset pricing model, which suggests that liquidity risk is a priced factor in asset pricing. Amihud, Mendelson and Pedersen (2005) classify sources of illiquidity, including exogenous transaction cost, exogenous trading horizons and time-varying trading costs, as well as clientele effects, etc.

Amihud and Mendelson (1986) argue that the most direct and fundamental source of illiquidity is the exogenous transaction cost, that the price discount of a stock due to illiquidity is the present value of the expected future stream of transaction costs, thus its return increases with regard to its illiquidity. They also factor in the exogenous trading horizons and time-varying trading costs, that a stock's additional return required to compensate an increase in spread is higher for lower-spread stocks than for higher-spread ones, and that illiquidity is time-varying prompting investors to require compensation. In addition, they consider clientele effects by studying a scenario in which diverse investors with different expected holding periods require different return levels. Their key argument is that, in equilibrium, liquid assets with lower transaction costs are more likely to be held by high-frequency traders while the illiquid assets with higher transaction costs are to be held by long-term investors, and that stocks with higher transaction costs are allocated to investors with longer investment horizons.

**2.3 Volatility of Liquidity Models**

Uncertain trading horizon contributes to liquidity risk (Amihud, Mendelson and Pedersen, 2005). Huang (2003) shows that in equilibrium, an illiquid stock with stochastic net return has a liquidity premium (over a liquid stock) that exceeds the expected transaction costs. Vayanos (2004) shows that liquidity premium is time-varying if an investor's liquidation risk (when to liquidate) is time-varying. In many cases, trading horizons are determined by investors' behavior, and they can tradeoff costs and benefits of delaying trades. These different endogenous trading horizons also contribute to liquidity risk. Constantinides (1986) and Vayanos (1998) find that, in certain cases, liquidity premium is much smaller than trading costs. Vayanos (1998) also shows in certain cases, transaction costs actually can make stock price higher, as trading costs induce investors to hold shares longer, which in turn raises their demand.

Time-varying liquidity affects volatility as it affects price (Amihud, Mendelson and Pedersen, 2005). This suggests a positive relation between returns and volatility of liquidity. If a stock's liquidity has a high volatility, there is a high probability of low liquidity at liquidation. Therefore, risk-averse investors are likely to hold stocks with low volatility of liquidity. Petkova, Akbas and Armstrong (2011) find that stocks with higher variability in liquidity command higher expected returns, suggesting that investors dislike volatility of liquidity due to possible liquidity downturns. On the other hand, Chordia, Subrahmanyam and Anshuman (2001) show that there is a strong and negative relation between the volatility of liquidity and expected returns. Pereira and Zhang (2010) find that stocks with higher volatility of liquidity tend to earn lower returns, and that investors with investment horizons sell their stocks during periods of high liquidity, revealing their preference due to possible liquidity upturns. Amiram et al. (2019) consolidate the abovementioned dispute by decomposing the total volatility into a jump component and a diffusion component. They observe a positive relation between the jump volatility and illiquidity and a negative relation between the diffusive volatility and illiquidity, which translates to different effects on liquidity risk premium for the two volatility components. Their findings are consistent with both bid-ask spread and Amihud (2002) illiquidity measures.

Andersen, Bollerslev, and Meddahie (2011) extend existing diffusive volatility models to the market microstructure noise, in which liquidity is factored in implicitly. Their results suggest that the detrimental impact of the microstructure noise on forecast accuracy can be substantial. As such, some alternative methods, including long memory and jump risk (again with liquidity being implicit), may be beneficial. Bollerslev and Todorov (2023) argue that the risk associated with market price jumps is priced differently from that with continuous price moves, and develop a series of model-free, short-time and risk-neutral variance expansions to delineate the importance of jumps in generating both price and variance risks. They find that simultaneous jumps in the price and the stochastic volatility, and/or jump intensity of the market command a sizeable risk premium. Similarly, Hu and Liu (2022) analyze the cross-section of index option returns by including volatility and jump risks, and argue that low option returns are primarily due to the pricing of market volatility risk. Hussain (2011) uses an EGARCH model to estimate the conditional return volatility. The model includes contemporaneous and lagged trading volume and bid-ask spreads for information arrival, and tests asymmetric reactions of return volatility in response to changes in volume and spreads.

**2.4 Models of Assets with Extreme Liquidity**

Most of the existing liquidity (and volatility of liquidity) models are not fully tested empirically on assets with extreme liquidity, primarily due to that the trading of a vast majority of liquid assets is regulated, as such does not exhibit statistically meaningful pattern of extreme liquidity. The rapid proliferation of crypto assets over the past ten years has created one of the largest unregulated alternative asset classes with excessive trading activities, and therefore offers an excellent empirical proving ground that enables researchers to model assets with extreme liquidity.

Using millisecond-level data for five major crypto assets and two crypto indices, Manahov (2021) shows that the traders facilitate extreme price movements and demand liquidity premium and that there is presence of herding behavior during up markets. Shen, Urquhart, and Wang (2022) examine intraday time series momentum for Bitcoin, and find that the return of the first half-hour trading positively predicts that of the second half-hour, suggesting an intraday trading momentum, and that the intraday momentum is driven by liquidity provision. Cong et al. (2023) investigate wash trading in crypto assets. They categorize 29 centralized crypto exchanges into two groups: a regulated one that resembles the transaction patterns of regular financial markets, and an unregulated one (two levels) that displays abnormal trading characteristics. They quantify that wash trades average more than 70% of the reported volumes for the unregulated exchanges, suggesting that crypto assets be modeled with purposely designed treatment on wash trades.

Having noticed possible autocorrelations in time series of crypto assets, Cortez, Rodríguez-García and Mongrutet (2021) compare the predictions on liquidity for crypto assets and fiat currencies with the ARMA and GARCH models, along with a machine learning algorithm (KNN). They find that the KNN approach is the best to capture liquidity for crypto assets in the short-term, while the ARMA models perform well for fiat currencies in developed markets, and the GARCH models do better for fiat currencies in emerging markets.

**2.5 Gaps in Liquidity Research**

At the theoretical level, the literature lacks models of return and volatility with explicit adjustment of liquidity. Although there exist studies on how liquidity interacts with return and volatility for both traditional assets (e.g., Amihud, Mendelson and Pedersen, 2005; Andersen,

Bollerslev, and Meddahie, 2011; Amiram et al., 2019) and alternative assets with extreme liquidity (e.g., Manahov, 2021; Shen, Urquhart, and Wang, 2022), none has been done on models to forecast the return and volatility of assets with explicit adjustment of liquidity. We aim to bridge this gap by providing models that explicitly adjust return and volatility with liquidity, which serves as our first motivation. As wash trades may artificially inflate the price and volume of crypto assets (Cong et al., 2023), we include specific treatments on wash trades (by removing/reducing possible wash trades) to further demonstrate the merits of our models.

At the methodological level, while the literature offers certain autoregressive models that address the liquidity risk (e.g., Hussain, 2011), these models are narrow in scope and seldom used in asset pricing. While there are autoregressive models for assets with extreme liquidity (e.g., Cortez, Rodríguez-García and Mongrutet, 2021), they are not effective as they are "standard" with no liquidity-adjustment. Some other studies model the liquidity effect implicitly through jumps with statistical models such as HAR (e.g., Corsi, 2009; Liu, Liu and Zhang, 2019). In this paper, we seek to develop a new set of liquidity-adjusted autoregressive models applicable to assets with extreme liquidity, along with improved predictability. This serves as our second motivation.

## 3. Liquidity-adjusted Volatility and Return and Liquidity Premium Beta

### 3.1 Liquidity-adjusted Volatility and Liquidity Measures

Imagine that a risk-averse investor had a choice between two assets, A and B, with identical expected return and volatility, yet A had a much higher trading volume/amount than B, the investor most likely would choose A, particularly if the investor's horizon were not indefinite. The intuitive reason for this choice is that the "perceived" volatility of A is lower than that of B. This choice is consistent with Amihud, Mendelson and Pedersen (2005). Using the tick-level trading data from the most dominant crypto asset exchange, Binance[4], we model this perceived and unobservable volatility as a minute-level liquidity-adjusted volatility $\sigma^{2\ell}_T$ at equilibrium for time-period $T$ (a 24-hour/1440-minute trading day) as follows:

$$\sigma^{2\ell}_T = \frac{1}{T}\sum_{t=1}^{T} \eta_T \frac{|r_t|/\overline{|r_t|}}{A_t/\overline{A_t}} (r_t - \bar{r}_t)^2 = \frac{1}{T}\sum_{t=1}^{T} \eta_T \ell_t (r_t - \bar{r}_t)^2 = \frac{1}{T}\sum_{t=1}^{T} \ell_T (r_t - \bar{r}_t)^2 \quad (2a)$$

$$\Rightarrow \sigma^{2\ell}_T = \frac{1}{T}\sum_{t=1}^{T} \left(\sqrt{\eta_T \frac{|r_t|/\overline{|r_t|}}{A_t/\overline{A_t}}} r_t - \sqrt{\eta_T \frac{|r_t|/\overline{|r_t|}}{A_t/\overline{A_t}}} \bar{r}_t\right)^2 = \frac{1}{T}\sum_{t=1}^{T} \left(\sqrt{\eta_T \ell_t} r_t - \sqrt{\eta_T \ell_t} \bar{r}_t\right)^2 = \frac{1}{T}\sum_{t=1}^{T} \left(\sqrt{\ell_T} r_t - \sqrt{\ell_T} \bar{r}_t\right)^2 \quad (2b)$$

*where:*

$$\ell_t = \frac{|r_t|/\overline{|r_t|}}{A_t/\overline{A_t}} \quad (2c)$$

$$\ell_T = \eta_T \frac{|r_t|/\overline{|r_t|}}{A_t/\overline{A_t}} = \eta_T \ell_t \quad (2d)$$

*subject to*:

$$\sum_{t=1}^{T} \eta_T \frac{|r_t|/\overline{|r_t|}}{A_t/\overline{A_t}} = \sum_{t=1}^{T} \eta_T \ell_t = \sum_{t=1}^{T} \ell_T = T \quad (2e)$$

$$\Rightarrow \eta_T = \frac{T}{\sum_{t=1}^{T} \frac{|r_t|/\overline{|r_t|}}{A_t/\overline{A_t}}} = \frac{T}{\sum_{t=1}^{T} \ell_t} \quad (2f)$$

*where $r_t$ is the observed return at minute t, $|r_t|$ is its absolute value, $\overline{|r_t|}$ is its arithmetic average in that day,*
*$A_t$ is the dollar (USDT) amount traded at minute t, $\overline{A_t}$ is its arithmetic average in that day,*
*$\eta_T$ is the daily normalization factor on day T and is a constant for day T,*
*T=1440 as there are 1440 minutes (24 hours) in a crypto asset trading day.*

Equation 2c states that $\ell_t$ is essentially a normalized (~1) minute-level version of Equation 1b of Amihud (2002), with both numerator and denominator of $\ell_t$ being normalized with regard to its respective parameters, the absolute return and trading amount ($|r_t|$ and $A_t$), and $\ell_T$ is a further normalized $\ell_t$ with a daily normalization factor $\eta_T$. The term of $\ell_t$ reflects the variability of a particular asset's intraday illiquidity, which is useful for intraday trading; the term of $\ell_T$ provides direct daily liquidity comparison between assets, which is useful for portfolio optimization.

### 3.2 Liquidity-adjusted Return and Liquidity Premium Beta

---

[4] https://coinmarketcap.com/rankings/exchanges/

Furthermore, in Equation 2a, $\sigma^2{}_T^\ell$ is the variance of an unobservable minute-level liquidity-adjusted return at equilibrium, $r_t^\ell$:

$$\sigma^2{}_T^\ell = \frac{1}{T}\sum_{t=1}^{T}\left(r_t^\ell - \overline{r_t^\ell}\right)^2 \tag{3}$$

where $r_t^\ell$ is the liquidity-adjusted return at minute $t$, $\overline{r_t^\ell}$ is its arithmetic average in that day.

Following the notion that there is a liquidity premium examined by the models of exogenous trading horizons and clientele effects (Amihud and Mendelson, 1986; Amihud, Mendelson and Pedersen, 2005), and by equating the right-hand side of Equations 2b and 3, we model an asset's observed return $r_t$ and its unobservable liquidity-adjusted counterpart at equilibrium $r_t^\ell$ (1st order approximation) as follows[5]:

$$r_t^\ell = \sqrt{\eta_T \frac{|r_t|/\overline{|r_t|}}{A_t/\overline{A_t}}} r_t = \sqrt{\eta_T \ell_t} r_t = \sqrt{\ell_T} r_t = \frac{1}{\beta_t^\ell} r_t \tag{4a}$$

$$\Rightarrow r_t = \beta_t^\ell r_t^\ell \tag{4b}$$

$$\Rightarrow \beta_t^\ell = \frac{r_t}{r_t^\ell} = \frac{1}{\sqrt{\eta_T \frac{|r_t|/\overline{|r_t|}}{A_t/\overline{A_t}}}} = \frac{1}{\sqrt{\eta_T \ell_t}} = \frac{1}{\sqrt{\ell_T}} \tag{4c}$$

where $r_t$ is observed return at time $t$, $r_t^\ell$ is the liquidity-adjusted return at time $t$,
$\beta_t^\ell$ is the liquidity premium factor at time $t$.

We structure $\beta_t^\ell$ in Equation 4c in such a way that it is a unitless liquidity measure and normalized (~1), and that when liquidity is high (amount is high), $r_t$ is greater than $r_t^\ell$, while liquidity is low (amount is low), $r_t$ is smaller than $r_t^\ell$. This reflects the intuition of a liquidity premium, that a perfectly liquid (frictionless, $A_t \to \infty$) stock generates no liquidity-adjusted return ($r_t^\ell \to 0$), and that a perfectly illiquid (infinite friction, $A_t \to 0$) stock earns an infinite liquidity-adjusted return ($r_t^\ell \to \infty$).[6]

---

[5] Note that although we define $r_t^\ell = \sqrt{\eta_T \ell_t} r_t$, in general $\overline{r_t^\ell} = \sqrt{\eta_T \ell_t} \overline{r_t}$ does not hold. However, its 1st order approximation is acceptable.
[6] Note that $\ell_t$ is zero if $r_t$ is zero in that minute, which makes $\beta_t^\ell$ infinite. To correct this singularity, we assign a random number two magnitudes ($10^{-2}$) smaller than $\overline{r_t}$ to each $r_t$ with a zero value.

We thus refer $\beta_t^\ell$ as the minute-level "liquidity premium Beta" given as:

$$\beta_t^\ell = \frac{1}{\sqrt{\eta_T \ell_t}} = \frac{1}{\sqrt{\ell_T}} \subset \begin{cases} > 1; high\ liquidity \\ = 1; equilibrium\ liquidity \\ < 1; low\ liquidity \end{cases} \qquad (5)$$

We then construct the unobservable daily (day-level) liquidity-adjusted return by aggregating the intraday (minute-level) liquidity-adjusted returns. The realized and unobservable daily liquidity-adjusted returns for time-period T are given as:

$$r_{TT} = (1 + r_t)^T - 1 \cong (1st\ order)\ \sum_{t=1}^T r_t \qquad (6)$$

$$r_{TT}^\ell = \left(1 + r_t^\ell\right)^T - 1 \cong (1st\ order)\ \sum_{t=1}^T r_t^\ell = \sum_{t=1}^T \sqrt{\eta_T \ell_t} r_t = \sum_{t=1}^T \sqrt{\ell_T} r_t = \sum_{t=1}^T \frac{r_t}{\beta_t^\ell} \qquad (7)$$

The daily liquidity premium Beta, $\beta_{TT}^\ell$, for time-period T is thus defined as follow[7]:

$$\beta_{TT}^\ell = |r_{TT}/r_{TT}^\ell| \subset \begin{cases} > 1; high\ daily\ liquidity \\ = 1; equilibrium\ daily\ liquidity \\ < 1; low\ daily\ liquidity \end{cases} \qquad (8)$$

The realized and unobservable daily minute-level (intraday) variance for time-period $T$ is:

$$\sigma^2{}_{TT}^\ell = T\sigma^2{}_T^\ell \qquad (9)$$

We will then fit $r_{TT}$ and $r_{TT}^\ell$ in the ARMA-GARCH/EGARCH models (see Section 5).

## 4. Data and Descriptive Statistics

### 4.1 Dataset

We first collect tick-level trading data of the top 15 crypto assets (prices measured with their trading pairs with Tether or USDT)[8] in terms of their market values from the Binance API.[9] Furthermore, in order to construct a portfolio with sufficient observations for back-tests, we narrow down our selection to ten crypto assets, all with at least 4.5 years of historical data (April 27, 2019 to January 1, 2024 with 1,711 trading days) for 3.5-year back-tests (April 26, 2020 to January 1, 2024, 1,346 trading days) with a 1-year (365 days) rolling window. The selected crypto assets include ADA, BNB, BTC, ETC, ETH, LINK, LTC, MATIC, XMR and XRP.

---

[7] Unlike $\beta_t^\ell$, it is possible that $r_{TT}/r_{TT}^\ell$ is negative, therefore $\beta_{TT}^\ell$, a positive value, is defined as $|r_{TT}/r_{TT}^\ell|$.
[8] USDT is a "stable coin" pegged to the US Dollar, and for our purpose of portfolio construction it is regarded as the "risk-free" asset with a 0% interest rate.
[9] The market value of each selected crypto asset counts for at least 0.25% of the overall crypto asset market cap. The total market value of the 15 crypto assets is more than 80% of the overall crypto asset market cap as of May 1, 2023.

We aggregate the tick-level data to construct and calculate minute-level (intraday) amount ($A_t$). As a treatment of wash trades (Cong et al., 2023), we divide $A_t$ into four quantiles (Q1 – Q4) of equal quantity, and reduce the quantity of Q3 (50 percentile) by a factor of 50%, and the quantity of Q4 (75 percentile) by a factor of 75%. From the minute-level data we further calculate daily amount $A_{TT}$, which is on average about 40% (see Subsection 4.2 for detailed statistics) of the untreated "raw" daily amount across the selected assets, or about 60% of the raw daily amount is regarded as being from wash trades. Since Cong et al. (2023) estimate that wash trades count for 46.47% of the total amount in Binance (an unregulated "tier 1" exchange), our treatment is actually more stringent, which makes our results robust.

We then construct and calculate minute-level (intraday) returns and variances, both regular ($r_t$ and $\sigma^2_T$) and liquidity-adjusted ($r_t^\ell$ and $\sigma^{2\ell}_T$). From the minute-level data we calculate their corresponding daily minute-level returns and variances, both regular ($r_{TT}$ and $\sigma^2_{TT}$) and liquidity-adjusted ($r_{TT}^\ell$ and $\sigma^{2\ell}_{TT}$), as well as the daily liquidity premium Beta $\beta^\ell_{TT}$, over the full sample period. We report the descriptive statistics of daily data for all ten crypto assets in Tables 1 to 4.

**4.2 Descriptive Statistics and Implications**

Table 1 provides the descriptive statistics of the daily regular return $r_{TT}$ (Panel A) and the daily liquidity-adjusted return $r_{TT}^\ell$ (Panel B). Panel A shows that the mean values of $r_{TT}$ range from 0.05% (XMR) to 0.56% (LINK), the minimum values range between -99.11% (BNB) and -38.53% (LTC), and the maximum values are between 44.98% (BTC) and 324% (ADA). The rather extreme returns on both ends reflect the fact that there is no regulatory oversight on crypto trading.

The measures for the daily liquidity-adjusted return $r_{TT}^{\ell}$ are summarized in Panel B: the mean values are between 0.62% (BNB) and 5.49% (LINK), the minimum values range from -99.76% (BNB) to -40.24% (BTC), and the maximum values are from 90.23% (BTC) to 836.28% (ADA).

From Panel B of Table 1, we observe that the mean values of $r_{TT}^{\ell}$ are numerically greater than that of $r_{TT}$ for all the assets (mostly by an order of magnitude), and the maximum values of $r_{TT}^{\ell}$ are numerically greater than that of $r_{TT}$ for all the assets except XRP, which seems to suggest that they would have a low daily liquidity ($|r_{TT}/r_{TT}^{\ell}| < 1$) on average. However, in Table 2, all ten assets have a mean daily liquidity premium Beta ($\beta_{TT}^{\ell}$) greater than 1, ranging from 1.20 (BTC) to 1.59 (ADA), indicating that all the assets actually have, on average, high daily liquidity.

The apparent contradiction can be explained by the heavily positive-skewed distribution of $\beta_{TT}^{\ell}$ for the full sample period: for all assets, the median of $\beta_{TT}^{\ell}$ is in general less than 1 with a value between 0.73 (BTC) and 0.91 (BNB) and is less than its mean, while the maximum exceeds 10.[10] The number of days with a $\beta_{TT}^{\ell}$ higher than its mean is between 376 (XRP) and 428 (MATIC) days (21.98% to 25.01% of 1,711 days). While the number of days with an extreme $\beta_{TT}^{\ell}$ (= 10) is only 44 (MATIC) to 61 (ADA and BNB) days (2.57% to 3.57% of 1,711 days), they carry 18.53% to 22.91% of the overall $\beta_{TT}^{\ell}$ load. A set of histograms (Figure 1) confirms that the $\beta_{TT}^{\ell}$ has a thin yet very long right tail (i.e., with some extremely large values) for all the assets.[11] Therefore, just a few days of heavy trading with extreme liquidity distorts the $\beta_{TT}^{\ell}$ distribution. Furthermore, the minimum $\beta^\ell_{TT}$ is almost always 0.00 (Table 2), thus there are also trading days with extremely low liquidity. Using a value of 0.10 as the lower-bound threshold, we identify 60 (XRP) to 132 (XMR) days out of 1,711 days (3.51% to 7.71%) with extremely low liquidity. That $\beta^\ell_{TT}$ has extreme values on both ends indicates that the distribution of $r^\ell_{TT}$ is wider than that of $r_{TT}$ for all assets.

---

[10] Note that we cap the $\beta_{TT}^{\ell}$ to 10 to avoid extreme large values (there would have been some), so that we will not have unreasonable asset allocations in liquidity-weighted portfolios (see more details in Section 6).
[11] There are spikes at $\beta_{TT}^{\ell}$ being equal to 10, and again we cap the $\beta_{TT}^{\ell}$ at 10 to avoid extremely large values.

The above observations can be further validated by the distribution of daily amount $A_{TT}$ (with treatments of wash trades). The histograms in Figure 2 illustrate that $A_{TT}$ is also highly positive-skewed (mean ≫ median) with long right tail. Panel A of Table 3 shows that aggregated amount in just 12 (ETC) to 104 (BTC) days (0.70% to 6.08% of 1,711 days) with the highest amount (in descending order) count for 25% of the overall trading value in the full sample period. Panel B of Table 3 shows that the $A_{TT}$ as a percentage of the raw daily amount (without treatments of wash trades) has an average value between 34.43% (XMR) and 46.54% (BTC), a lot more stringent than the estimate of Cong et al. (2023) that wash trades count 46.47% of the trades in Binance.

Tables 4 shows that, for all the assets, the difference between the minimum values of regular and liquidity-adjusted daily minute-level (intraday) variances ($\sigma^2_{TT}$ vs. $\sigma^{2\ell}_{TT}$) is not numerically significant (all very close to 0%). On the other hand, the liquidity-adjusted variance $\sigma^{2\ell}_{TT}$ has a higher maximum, indicating that its distribution is wider than that of $\sigma^2_{TT}$, which is consistent with the distribution of $r^\ell_{TT}$ being wider than that of $r_{TT}$. Therefore, liquidity-adjusted return and variance ($r^\ell_{TT}$-$\sigma^{2\ell}_{TT}$) have a wider 2-dimensional distribution than the regular return and variance ($r_{TT}$-$\sigma^2_{TT}$). In addition, the highly asymmetric distribution of $\beta^\ell_{TT}$ transfers the distribution of $r_{TT}$-$\sigma^2_{TT}$ into a wider and highly asymmetric distribution of $r^\ell_{TT}$-$\sigma^{2\ell}_{TT}$, which is "tighter" (than that of $r_{TT}$-$\sigma^2_{TT}$) when liquidity is high ($\beta^\ell_{TT} > 1$), but "looser" when liquidity is low ($\beta^\ell_{TT} < 1$). A set of visualizations in Figure 3 illustrate that the distributions of $r^\ell_{TT}$-$\sigma^{2\ell}_{TT}$ are indeed wider than that of $r_{TT}$-$\sigma^2{}_{TT}$, tighter with high liquidity (low range of return-volatility space) and looser with low liquidity (high range of return-volatility space).[12]

As such, the wide and positive-skewed distribution of liquidity ($\beta^\ell_{TT}$) indeed brings in asymmetric (non-normal) information from a third dimension of liquidity, in addition to the information already embedded in the "standard" daily regular return and variance. Therefore, the seemingly 2-dimensional distribution of the liquidity-adjusted return and volatility ($r^\ell_{TT}$-$\sigma^{2\ell}{}_{TT}$) is actually a "flattened" 3-dimensional (regular return, volatility, and liquidity: $r_{TT}$-$\sigma^2{}_{TT}$-$\beta^\ell_{TT}$) information ensemble. A set of scatter plots in Figure 4 illustrate the 3-dimensional distribution of return-volatility-liquidity ($r_{TT}$-$\sigma^2{}_{TT}$-$\beta^\ell_{TT}$), which visualizes the additional information from the third dimension of liquidity (distribution of $\beta^\ell_{TT}$).[13]

In addition, we investigate on when the extreme liquidity occurs over the full sample period for each asset through the visualization of $A_{TT}$ (Figure 5). We find that days with high amounts tend to cluster (so do the days with low amounts). This means that the amount, thus the liquidity, demonstrates a "stickiness" property (time series). Furthermore, we observe that extreme liquidity, either high or low, occurs at about the same time for all assets (cross-sectional). Therefore, when extreme liquidity occurs for all the assets, there is also extreme liquidity on the portfolio level.

## 5. Liquidity-Adjusted ARMA-GARCH/EGARCH Models

In this section we present a liquidity-adjusted ARMA-GARCH/EGARCH construct for the modeling of univariate condition return for individual assets. We theoretically demonstrate that the wide, positive-skewed and sticky distribution of liquidity improves the predictability of these models over their traditional counterparts.

---

[12] In Figure 3, $\sigma_{TT}$ is used instead of $\sigma^2{}_{TT}$, which is standard in return-volatility visualization.
[13] Again, in Figure 4, $\sigma_{TT}$ is used instead of $\sigma^2{}_{TT}$, which is standard in return-volatility visualization.

A set of Adam-Fuller tests (Table 5) confirms that both $r_{TT}$ and $r_{TT}^{\ell}$ for all ten crypto assets are stationary time series over the entire 1,711-day period. We then apply the standard univariate ARMA(*p,q*)-GARCH/EGARCH(1,1) model on both $r_{TT}$ and $r_{TT}^{\ell}$ for each asset to estimate their conditional returns and variances for the in-sample observations (from Apr 26, 2020 to January 1, 2024 with 1,346 trading days) with a 365-day rolling window. The time series ARMA-GARCH/EGARCH construct is given in Equations 10.[14]

In the ARMA stage, we use the Akaike Information Criterion (AIC) to select the best-fit values of *p* and *q* (p, q ≤ 4) for each rolling window (Equation 12a). In the GARCH/EGARCH stage, we adopt the (1,1) specification to reduce computation penalty and use the AIC criteria to select between a GARCH (Equation 10b) specification and an EGARCH specification (Equation 10c) for each in-sample day *t*:

$$ARMA(p,q): r_t = \delta + \sum_{i=1}^{p} \phi_i r_{t-i} - \sum_{j=1}^{q} \theta_j \epsilon_{t-j} + \epsilon_t \quad (10a)$$

$$GARCH(1,1): \sigma_t^2 = \omega + a\epsilon_{t-1}^2 + b\sigma_{t-1}^2 \quad (10b)$$

$$EGARCH(1,1): \log \sigma_t^2 = \omega + bg(Z_{t-1}) + a \log \sigma_{t-1}^2 \quad (10c)$$

$$\text{where } g(Z_t) = \theta Z_t + \lambda(|Z_t| - E(|Z_t|)); \ Z_t \sim N(0,1)$$

We then apply Equations 10 to produce the one-period (*t+1*) forecasted conditional mean return ($\hat{\mu}_{t+1}^{arga}$) for each rolling window for both $r_{TT}$ and $r_{TT}^{\ell}$, and report the forecast accuracy of conditional returns in terms of Root Mean Squared Error (RMSE) in Table 5.[15]

---

[14] From this point on, the subscript *t* refers to a point in time with a daily interval, i.e., day *t*
[15] For the rest of the paper, the "current timestamp" is end of day *t*, thus a variable with a *t* subscript is "realized" (either a direct observation or a calculated value from direct observations), while a variable with a *t+1* subscript is a one-period forecasted value. We also use accent mark "$\bar{v}$" to represent a mean variable, accent mark "$\hat{v}$" to represent a forecasted variable, and no accent mark "$v$" to represent a realized variable.

While we apply the standard univariate ARMA-GARCH/EGARCH model of Equations 10 to both $r_{TT}$ and $r_{TT}^{\ell}$, a liquidity-adjusted ARMA-GARCH/EGARCH model is warranted to capture the impact of the additional dimension of liquidity embedded in $r_{TT}^{\ell}$. We thus derive the generalized liquidity-adjusted ARMA-GARCH/EGARCH at extreme liquidity as follows[16]:

$$ARMA: r_t^{\ell} = \sum_{i=1}^{p} \phi_i r_{t-1}^{\ell} - \sum_{j=1}^{q} \theta_j \epsilon_{t-j}^{\ell} + \epsilon_t^{\ell} \tag{11a}$$

$$GARCH(1,1): \sigma_t^{\ell 2} = \omega + a\epsilon_{t-1}^{\ell\,2} + b\sigma_{t-1}^{\ell\,2} + \omega \frac{1-\beta_{t-1}^{\ell\,2}}{\beta_{t-1}^{\ell\,2}} \tag{11b}$$

$$EGARCH(1,1): \log \sigma_t^{\ell 2} = \omega + b g(Z_{t-1}) + a \log \sigma_{t-1}^{\ell\,2} + (a-1)\log \beta_{t-1}^{\ell\,2} \tag{11c}$$

where:

$$g(Z_t) = \theta Z_t + \lambda(|Z_t| - E(|Z_t|)); \; Z_t \sim N(0,1)$$

The GARCH(1,1)/EGARCH(1,1) specifications for volatility of error term for the liquidity-adjusted return $r_t^{\ell}$ (Equations 11b and 11c) have an extra liquidity term ($\omega \frac{1-\beta_{t-1}^{\ell\,2}}{\beta_{t-1}^{\ell\,2}}$ and $(a-1)\log \beta_{t-1}^{\ell\,2}$, respectively), compared to that for the regular return $r_t$ (Equations 10b and 10c). Since the extra liquidity terms can be either greater or less than 0 (depending on whether $\beta_{t-1}^{\ell}$ is greater or less than 1), no additional treatment on sign asymmetry is necessary. The error and volatility terms of the standard models of Equations 10 are connected to that of the liquidity-adjusted models of Equations 11 as follows:

$$\epsilon_t^{\ell} = \frac{\epsilon_t}{\beta_t^{\ell}} \tag{12a}$$

$$\sigma_t^{\ell} = \frac{\sigma_t}{\beta_t^{\ell}} \tag{12b}$$

where:

$$\epsilon_t | \Psi_{t-1} = \sigma_t z_t; z_t \sim \mathcal{N}(0,1)$$
$$\epsilon_t^{\ell} | \Psi_{t-1} = \sigma_t^{\ell} z_t; z_t \sim \mathcal{N}(0,1)$$

---

[16] The details of deriving the liquidity-adjusted ARMA-GARCH/EGARCH model are given in Appendix 1.

Therefore, with extreme values of $\beta_t^\ell$, the liquidity-adjusted ARMA-GARCH/EGARCH model of Equations 11 has an error term $\epsilon_t^\ell$ equal to $\epsilon_t$ (of the standard model) scaled by $\beta_t^\ell$ (Equation 12a). As such, when $\beta_t^\ell > 1$, $\epsilon_t^\ell < \epsilon_t$, when $\beta_t^\ell < 1$, $\epsilon_t^\ell > \epsilon_t$. Conceptually, this means that when liquidity is high, the liquidity-adjusted model offers better predictability for the liquidity-adjusted return $r_t^\ell$ than the regular model for the regular return $r_t$, while when liquidity is low, the opposite is true. In addition, the conditional volatility (standard deviation) $\sigma_t^\ell$ of the error term $\epsilon_t^\ell$ is equal to $\sigma_t$ (of the standard model) scaled by $\beta_t^\ell$ (Equation 12b): when $\beta_t^\ell > 1$, $\sigma_t^\ell < \sigma_t$, when $\beta_t^\ell < 1$, $\sigma_t^\ell > \sigma_t$. Conceptually, this means that when liquidity is high, the liquidity-adjusted model produces lower volatility of error term for the liquidity-adjusted return $r_t^\ell$, compared to the standard model for the regular return $r_t$. When liquidity is low the opposite is true. The scaling effect of $\beta_t^\ell$ on $\sigma_t^\ell$ is the same as that on $\epsilon_t^\ell$. Therefore Equation 14 provides the perspective that when liquidity is high ($\beta_{TT}^\ell > 1$), a "tighter" liquidity-adjusted distribution of $r_{TT}^\ell$-$\sigma^2{}_{TT}^\ell$ (compared to that of $r_{TT}$-$\sigma^2{}_{TT}$) affords the liquidity-adjusted ARMA-GARCH/EGARCH better predictability with lower error and volatility ($\epsilon_t^\ell = \frac{\epsilon_t}{\beta_t^\ell}; \sigma_t^\ell = \frac{\sigma_t}{\beta_t^\ell}$), compared to its standard counterpart.

In Table 5, the majority (9 out of 10) of the forecasted liquidity-adjusted returns have higher RMSE values (lower accuracy), compared to those for the regular returns, which seems to suggest that the ARMA-GARCH/EGARCH offers better predictability for the regular returns on average. This is again due to the highly distorted distribution of $\beta_{TT}^\ell$, that a large majority of trading days have low liquidity.[17] In Section 6 we will provide evidence that the LAMV has a clear performance improvement over the TMV because of the better ARMA-GARCH/EGARCH predictability for just a few days with extremely high liquidity.

---

[17] Table 2 shows that number of days with $\beta_{TT}^\ell$ being greater than 1.0 is between 525 (BTC) and 749 (BNB), which is only between 30.68% and 43.78% of the total number of trading days, and that there are more days with $\beta_{TT}^\ell$ being less than 0.1 (with no lower limit) than days with $\beta_{TT}^\ell$ being capped at 10 for all assets.

## 6. Empirical Tests on Liquidity-Adjusted Autoregressive Models

In this section we provide empirically evidence that the liquidity-adjusted ARMA-GARCH/EGARCH models of Section 5 offer improved predictability from their traditional counterparts, through the performance comparisons of a series of MV portfolios taking the forecasts of these models (conditional asset returns) as inputs.

### 6.1 Benchmark Portfolios

With a rolling window of 365 days over an out-of-sample period with 1,376 days, we construct four benchmark portfolios of the ten crypto assets (the risk-free asset USDT has a zero weight) for portfolio performance comparisons:

1. Portfolio 1: An equal-weight portfolio, all assets have an equal weight (10%).
2. Portfolio 2: A market (equilibrium) portfolio, each asset has a portfolio weight proportional to its market weight (in amount).
3. Portfolio 3: A liquidity-weight portfolio, each asset has a portfolio weight proportional to its daily liquidity premium Beta $(\propto \beta_{TT}^{\ell})$. This portfolio is suitable for investors with shorter investment-horizon that take advantage of low transaction costs.
4. Portfolio 4: An inverse-liquidity-weight portfolio, each asset has a portfolio weight inversely proportional to its daily liquidity premium Beta $(\propto 1/\beta_{TT}^{\ell})$. This portfolio is suitable for investors with longer investment-horizon that seek a liquidity premium.

### 6.2 Standard MV Portfolios

We then construct two MV portfolios: traditional and liquidity-adjusted. The standard daily-optimized MV in a time-series construct can be analytically expressed as the following quadratic programming problem with constraints:

$$\max_{W_t} \left( \bar{\mu}_t W_t^{Tr} - \frac{\lambda_t}{2} W_t^{Tr} \bar{\Sigma}_t W_t \right); Tr \text{ stands for Transposed} \tag{13}$$

subject to:

$$\sum_i^N w_t^i = 1; \ i = USDT, ADA, BNB, BTC, ETC, ETH, LINK, LTC, MATIC, XMR, XRP; \ N = 11$$

$w_t^i \geq 0$; *long-only*

$w_t^{USDT} \leq 1$

$w_t^i \leq 0.300 \ (3 \times equal \ weight); \ i \neq USDT$

where:

$$\lambda_t = \frac{r_{t_{mkt}}^P - r_{t_{risk\_free}}}{\sigma_{t\,mkt}^2} = \frac{r_{t_{mkt}}^P - r_{t_{USDT}}}{\sigma_{t\,mkt}^2} = \frac{r_{t_{mkt}}^P}{\sigma_{t\,mkt}^2}$$

$r_{t_{mkt}}^P$ is the return of the market or equilibrium portfolio on day t, $\sigma_{t\,mkt}^2$ is the variance;
$r_{t_{risk\_free}}$ is the return of a risk-free asset, and USDT is regarded as risk-free with 0% return

In the standard MV construct of Equation 13, $\bar{\mu}_t$ is the mean return vector of the ten-asset portfolio over a 365-day rolling window ending on day *t*, and $\bar{\Sigma}_t$ is the covariance matrix of daily returns of the constituent assets in that rolling window. Both $\bar{\mu}_t$ and $\bar{\Sigma}_t$ are derived from available information up to day *t*. In addition, $W_t$ is the portfolio weight vector to be optimized for day *t*. The daily MV portfolios are:

5. Portfolio 5: A TMV portfolio; $\bar{\mu}_t$ is the mean vector of $r_{TT}$'s over the rolling window ending on day *t*, or $\bar{\mu}_{r_t}$; $\bar{\Sigma}_t$ is the covariance matrix of $r_{TT}$'s for the rolling window, or $\bar{\Sigma}_{r_t}$.

6. Portfolio 6: A LAMV portfolio; $\bar{\mu}_t$ is the mean vector of $r_{TT}^{\ell}$'s over the rolling window ending on day *t*, or $\bar{\mu}_{r_t^{\ell}}$; $\bar{\Sigma}_t$ is the covariance matrix of $r_{TT}^{\ell}$'s for the rolling window, or $\bar{\Sigma}_{r_t^{\ell}}$.

### 6.3 ARMA-GARCH/EGARCH-enhanced MV Portfolios

To further improve the performance of Portfolios 5 and 6, we need to forecast the daily return vector. We construct two MV portfolios with forecasted daily return vectors (Portfolios 7 and 8).

To construct two ARMA-GARCH/EGARCH-enhanced MV portfolios with forecasted daily return vector we rewrite Equation 22 by retaining $\bar{\Sigma}_t$ and replacing $\bar{\mu}_t$ with the ARMA-GARCH/EGARCH forecasted return vector, $\hat{\mu}_{t+1}^{arga}$. The portfolios are constructed as:

$$\max_{W_t} \left( \hat{\mu}_{t+1}^{arga} W_t^{Tr} - \frac{\lambda_t}{2} W_t^{Tr} \bar{\Sigma}_t W_t \right) \quad (14)$$

All constraints for Equation 14 are the same as that in Equation 13. The ARMA-GARCH/EGARCH-enhanced MV portfolios are:

7. Portfolio 7: An ARMA-GARCH/EGARCH-enhanced TMV portfolio; $\hat{\mu}_{t+1}^{arga}$ is the return vector of ARMA-GARCH/EGARCH forecasted $r_{TT}$ values for day $t+1$, $\hat{\mu}_{r_{t+1}}^{arga}$; $\bar{\Sigma}_t$ is the covariance matrix of $r_{TT}$'s for the rolling window, or $\bar{\Sigma}_{r_t}$.

8. Portfolio 8: An ARMA-GARCH/EGARCH-enhanced LAMV portfolio; $\hat{\mu}_{t+1}^{arga}$ is the return vector of ARMA-GARCH/EGARCH forecasted $r_{TT}^{\ell}$ values for day $t+1$, $\hat{\mu}_{r_{t+1}^{\ell}}^{arga}$; $\bar{\Sigma}_t$ is the covariance matrix of $r_{TT}^{\ell}$'s for the rolling window, or $\bar{\Sigma}_{r_t^{\ell}}$.

## 6.4 Empirical Tests: Performance Comparisons of Portfolios with Wash Trades Removed

The performance of all portfolios is calculated as follow:

$$r_{t+1}^P = \mu_{t+1} W_t^{Tr} \quad (15a)$$

$$\sigma_{t+1}^{2P} = W_t \Sigma_{t+1}^{TT} W_t^{Tr} \quad (15b)$$

$$\sigma_{t+1}^P = \sqrt{\sigma_{t+1}^{2P}} \quad (15c)$$

Where, $r_{t+1}^P$, and $\sigma_{t+1}^{2P}$, $\sigma_{t+1}^P$ are the realized daily portfolio return, variance and standard deviation on day $t+1$;
$\mu_{t+1}$ is the realized daily return vector of constituent assets ($r_{t+1}$'s) on day $t+1$;
$W_t$ is the optimized portfolio weight on day $t$. All variables are annualized.

We use the annualized Sharpe ratio to compare the performance between portfolios:

$$SharpeR_{t+1} = \frac{r_{t+1}^P - r_{t+1 risk-free}}{\sigma_{t+1}^P} = \frac{r_{t+1}^P - r_{t+1 USDT}}{\sigma_{t+1}^P} = \frac{r_{t+1}^P}{\sigma_{t+1}^P} \quad (16)$$

Panel A of Table 6 gives the performance summaries of all 8 portfolios. Among the benchmark portfolios (Portfolios 1 – 4), Portfolio 1 (equal-weight portfolio) has the overall best performance, with the Sharpe ratio (1.52). We arrange the four MV portfolios (Portfolios 5 to 8) in such a way in Panel A of Table 6: the TMV portfolios with incremental forecast enhancement are on the left (Portfolios 5 and 7), while their corresponding LAMV portfolios are on the right (Portfolios 6 and 8). That way, we can easily observe and compare the incremental performance improvement after applying each type of enhancement methodology vertically within the TMV or LAMV, while at the same time conveniently compare the portfolio performances between the TMV and LAMV after applying one specific incremental enhancement methodology horizontally.

It is interesting that while the LAMV portfolios demonstrate a clear incremental improvement with each enhancement, the TMV portfolios actually exhibit a performance deterioration. In fact, among the TMV portfolios, the standard MV portfolio (Portfolio 5) without any forecast enhancement has better performance (Sharpe ratio of 1.43). With the mean return vector $\bar{\mu}_{r_t}$ being replaced by $\hat{\mu}^{arga}_{r_{t+1}}$, Portfolio 7 has a drastic performance decline from Portfolio 5, with the Sharpe ratio dropping to 0.81, the lowest among all TMV portfolios. This result provides strong empirical evidence that without incorporating the information on liquidity, the conditional return (non-liquidity-adjusted and therefore with jumps) is not well-behaved from the perspective of autoregression, thus the ARMA-GARCH/EGARCH is not effective and actually models $r_{TT}$ with worse perspective for day $t+1$ than just averaging $r_{TT}$'s with a rolling window. This finding is consistent with the literature, such as the HAR that averages realized volatility (Corsi, 2009).

The LAMV portfolios offer a stark contrast in that each forecast enhancement on the liquidity-adjusted variables induces performance improvement by significant numerical margins. Portfolio 6 is the standard LAMV portfolio with a Sharpe ratio of 1.02. Portfolio 8 incorporates $\hat{\mu}^{arga}_{r^{\ell}_{t+1}}$ and has a greatly improved performance over Portfolio 6, raising the Sharpe ratio to 1.51. This result provides clear empirical support to our hypothesis, that liquidity adjustment on jumps (extreme liquidity) restores the well-behaved autoregressive characteristics to $r^{\ell}_{TT}$, making the ARMA-GARCH/EGARCH effective again in modeling it. The LAMV Portfolio 8 has the best overall performance amongst the 8 portfolios.

We now compare the TMV and LAMV portfolios with the same forecast enhancement. The TMV Portfolio 5 has a higher Sharpe ratio than the LAMV Portfolio 6 (1.43 vs. 1.02). This means that when wash trades are excluded completely, the LAMV does not offer performance improvement over the TMV without forecast enhancement (In Subsection 6.5 we will show that with wash trades retained, LAMV has better performance).[18] For the ARMA-GARCH/EGARCH-enhanced portfolios (Portfolios 7 and 8), the LAMV Portfolio 8 has clearly superior performance over the TMV Portfolio 7 with much higher Sharpe ratio (1.51 vs. 0.81). As the only difference between the TMV and LAMV models is that the latter models conditional return with liquidity adjustment, we can conclude that its superior performance is a sole outcome of incorporating treatments of liquidity risk.

In summary, based on the comparisons of portfolio performance, the liquidity-adjusted ARMA-GARCH/EGARCH models are highly effective in modeling the daily liquidity-adjusted return $r_{TT}^{\ell}$. However, the standard ARMA-GARCH/EGARCH models are not effective in modeling the daily regular return $r_{TT}$, mainly because of the models lack information on liquidity. The visual performance comparisons of the eight portfolios are given in Figure 6.

**6.5 Robustness Tests: Performance Comparisons of Portfolios with Wash Trades Retained**

---

[18] In addition, in Subsection 6.6 we will discuss the effects of number of days with extreme liquidity on portfolio performance. Our goal is to demonstrate that the predictability of the liquidity-adjusted ARMA-GARCH/EGARCH models improves with increased number of days with extreme liquidity. We do not aim to produce the best-performed MV portfolios, we just use the MV portfolios to provide empirical evidence to our hypothesis.

The performance comparisons of Portfolios 1 to 8 in Subsection 6.4 provide empirical evidence that the liquidity-adjusted ARMA-GARCH/EGACH models, with wash trades removed, offer superior predictability to their traditional counterparts. To check the robustness of the models, we conduct additional tests with different levels of treatment on wash trading. Specifically, we rebuild the models based on raw trading amounts with no treatment on wash trading, upon which we reconstruct Portfolios 1 to 8 (portfolios with wash trades retained) and compare their performances.

Panel B of Table 6 summarizes the performances of the revised Portfolios 1 to 8 with no treatments on wash trading. Again, for each forecast enhancement, the TMV portfolios exhibit a performance deterioration, while the LAMV portfolios demonstrate a clear incremental improvement. Specifically, the TMV Portfolio 5 without any forecast enhancement has better performance (Sharpe ratio of 1.39) than the TMV Portfolio 7 enhanced with the traditional ARMA-GARCH/EGARCH forecast (Sharpe ratio of 0.85). On the other hand, the standard LAMV Portfolio 6 has a Sharpe ratio of 1.54, which is greatly improved by the forecast-enhanced LAMV Portfolio 8 to a value of 2.06. These results, again, provide a clear empirical support to our hypothesis that the adjustment on jumps (extreme liquidity) restores the well-behaved autoregressive characteristics to the liquidity-adjusted return $r_{TT}^{\ell}$, but at a different level of treatment on wash trades (specially, no treatment at all), indicating that our liquidity-adjusted ARMA-GARCH/EGARCH models are robust. The visual performance comparisons of the eight portfolios (wash trades retained) are given in Figure 7.

**6.6 Response of Liquidity-Adjusted ARMA-GARCH/EGARCH to Liquidity Level**

In this section we compare the performances of portfolios with and without treatment of wash trades, but otherwise identical in terms of incremental forecast enhancement. Through the comparisons, we validate that the liquidity-adjusted ARMA-GARCH/EGARCH models have rationally response to different levels of liquidity, and therefore are robust.

From Panel C of Table 6, for the benchmark portfolios with wash trades removed, Portfolios 3 (liquidity-weighted) and 4 (inverse-liquidity-weighted) have rather different performances, with Portfolio 3 having a clear advantage (Sharpe ratio of 1.43 vs. 0.97). This means that, with wash trades removed (thus fewer number of days with extreme liquidity, as well as lower overall level of liquidity)[19], shorter-term investors taking advantage of low transaction costs do better than long-term investors that seek high liquidity premium. However, for the portfolios with wash trades retained, the performances of Portfolios 3 and 4 are much closer than that of their wash-trades-removed equivalents (Sharpe ratio of 1.20 vs. 1.13). This means that, when there are greater number of days with extreme liquidity and higher level of overall liquidity, long-term investors get higher liquidity premium, while shorter-term investors benefit less from lower transaction costs. The reason is that, we conjecture, at least a portion of the wash trades are initiated by liquidity providers (automated exchange market makers and high frequency traders), which actually improve the pricing efficiency and thus provide long-term price support.

We observe that, in general, the performances of TMV portfolios with wash trades removed and retained are similar, which is expected as the regular return contains no liquidity information. On the other hand, the performance of LAMV portfolios with wash trades retained sees improvement from that with wash trades removed. Portfolio 8 (wash trades retained) has a greatly improved performance over Portfolio 8 (wash trades removed) with the Sharpe ratio rising to 2.06 from 1.51. The results indicate that the predictability of the liquidity-adjusted ARMA-GARCH/EGARCH models improves proportionally to the number of days with extreme liquidity and the overall level of liquidity, which again provides validation that our models are robust in that they respond rationally to different levels of liquidity (i.e., treatment on wash trading).

---

[19] We have full data on all liquidity measures, liquidity-adjusted return and volatility, and forecasted values with wash trades retained. The data is available upon request.

In addition, when comparing the performances of TMV and LAMV portfolios with the same forecast enhancement at the same level of treatment on wash trades, we notice that, with the wash trades retained, the Sharpe ratio (1.54) of the standard LAMV Portfolio 6 is higher than that (1.39) of the standard TMV Portfolio 5, whereas with wash trades removed the opposite is true (see discussion in Subsection 6.4). This suggests that there seems to be a threshold in number of days, that once the number of days with extreme liquidity is above the threshold, there would be sufficient number of adjacent days with extreme liquidity (liquidity is sticky) that makes the liquidity-adjusted return (and portfolio covariance) more reflective on the momentum of asset price, resulting in improved portfolio optimization. From the perspective of investors, they may actually benefit from wash trades designed to profit their initiators, who may unwillingly provide predictable momentum that attracts herds. As such, wash trades may include not only actions from their manipulative initiators, but also reactions from the herds. This would be another explanation on the herd trading during up markets (Manahov, 2021) as high volume provides predictable momentum (Shen, Urquhart, and Wang, 2022). Combining this with that they may provide long-term price support, wash trades may actually bring benefits to some investors.

## 7. Conclusions

In this paper, we hypothesize that asset price jumps are manifestations of extreme liquidity, and if we adjust the return and volatility of assets (with extreme liquidity) with liquidity, their well-behaved autoregressive properties would be restored, and the AMRA-GARCH/EGARCH would become effective again in modeling them. We develop liquidity measures on both minute and daily levels with treatments of wash trades, as well as liquidity-adjust return and volatility. We thoroughly analyze the distribution of liquidity-adjusted return and volatility of assets with extreme liquidity. We further examine how the distribution of liquidity-adjusted return and volatility affects the predictability of a set of liquidity-adjusted ARMA-GARCH/EGARCH models. To empirically test whether the liquidity-adjusted autoregressive models provide improved predictability, we construct a series of MV portfolios of selected crypto assets to compare their performances. We choose crypto assets because they are the most visible class of assets with extreme liquidity.

We find that the distribution of daily liquidity premium Beta (and that of the daily trading amount) is heavily positive-skewed with a thin yet very long right tail for all the crypto assets. Its distribution is wide with both extremely high and low values, and is "sticky" with clustering days of extreme liquidity (either high or low). Because of its wide distribution, the liquidity-adjusted return and volatility have wider 2-dimensional distribution than the regular return and variance. The clustering of extreme liquidity "shrinks" the distribution when liquidity is high, and "enlarges" it when liquidity is low. Therefore, the 2-dimensional distribution of liquidity-adjusted return and volatility is the projection of the 3-dimensional distribution of regular return-volatility-liquidity.

We then develop a set of liquidity-adjusted ARMA-GARCH/EGARCH to model the daily (conditional) liquidity-adjusted return for individual assets, which better predictability at extreme liquidity. We provide empirical evidence by comparing the performances of the LAMV and TMV portfolios. Furthermore, we validate the robustness of the liquidity-adjusted autoregressive models by repeating the above procedure without treatment of wash trading.

In addition, we find that the predictability of the liquidity-adjusted ARMA-GARCH/EGARCH models improves proportionally to the number of days with extreme liquidity and the overall level of liquidity. Also, we demonstrate that if the number of days with extreme liquidity is above a certain threshold, the liquidity-adjusted return and variance are more reflective on the momentum of asset price. We suggest that wash trades provide long-term price support and provide predictable momentum, therefore wash trades may be beneficial for investors.

In conclusion, our hypothesis is accepted with confidence. Our liquidity-adjusted models provide enhanced predictability for assets with extreme liquidity, and provide a viable and robust alternative for jump and statistical models (e.g., HAR), and can be utilized to measure and analyze the effect of liquidity for other asset classes with high liquidity risk.

**Appendix – Deriving Liquidity-Adjusted ARMA-GARCH/EGARCH**

The standard univariate ARMA($p,q$)-GARCH/EGARCH(1,1) specification is given as follows:

$ARMA(p,q): r_t = \delta + \sum_{i=1}^{p} \phi_i r_{t-i} - \sum_{j=1}^{q} \theta_j \epsilon_{t-j} + \epsilon_t$ (A1a)

$GARCH(1,1): \sigma_t^2 = \omega + a\epsilon_{t-1}^2 + b\sigma_{t-1}^2$ (A1b)

$EGARCH(1,1): \log \sigma_t^2 = \omega + b g(Z_{t-1}) + a \log \sigma_{t-1}^2$ (A1c)

where $g(Z_t) = \theta Z_t + \lambda(|Z_t| - E(|Z_t|))$; $Z_t \sim N(0,1)$

In order to demonstrate how the daily liquidity-adjusted return $r_{TT}^\ell$ responds to the ARMA-GARCH/EGARCH models, we start from an ARMA(1,0)-GARCH(1,1)/EGARCH (1,1) specification that links the regular observable return $r_{TT}$ to the liquidity-adjusted return $r_{TT}^\ell$. The standard ARMA(1,0) specification is given as:[20]

$r_t = \phi r_{t-1} + \epsilon_t$ (A2a)

$\Rightarrow \beta_t^\ell r_t^\ell = \phi \beta_{t-1}^\ell r_{t-1}^\ell + \epsilon_t$ (A2b)[21]

$\Rightarrow r_t^\ell = \phi \frac{\beta_{t-1}^\ell}{\beta_t^\ell} r_{t-1}^\ell + \frac{\epsilon_t}{\beta_t^\ell} = \phi_\ell r_{t-1}^\ell + \frac{\epsilon_t}{\beta_t^\ell}$ (A2c)

where: $\phi_\ell = \phi \frac{\beta_{t-1}^\ell}{\beta_t^\ell}$ (A2d)

We have established that the distribution of extreme values of $\beta_{TT}^\ell$ ($\beta_t^\ell$ in Equations A1) tend to cluster, therefore when $\beta_t^\ell \gg 1$ or $\beta_t^\ell \ll 1$, $\frac{\beta_{t-1}^\ell}{\beta_t^\ell} \approx 1$ and $\phi_\ell \approx \phi$, we thus have the following liquidity-adjusted ARMA(1,0) specification:

$r_t^\ell = \phi_\ell r_{t-1}^\ell + \epsilon_t^\ell \approx \phi r_{t-1}^\ell + \epsilon_t^\ell$ (A3a)

where: $\epsilon_t^\ell = \frac{\epsilon_t}{\beta_t^\ell}$ (A3b)

Therefore, with extreme values of $\beta_t^\ell$, the ARMA(1,0) specification of the liquidity-adjusted return (Equation A3a) has the same autoregressive coefficient ϕ as that of the regular return (Equation A2a), with an error term $\epsilon_t^\ell$ equal to $\epsilon_t$ being scaled by $\beta_t^\ell$ (Equation A3b). As such, when $\beta_t^\ell > 1$, $\epsilon_t^\ell < \epsilon_t$, and when $\beta_t^\ell < 1$, $\epsilon_t^\ell > \epsilon_t$. Conceptually, this means that when liquidity is high, the ARMA model offers better predictability for $r_t^\ell$ than $r_t$, while when liquidity is low, the opposite is true. We model the error term in the standard ARMA of Equation A2a with the GARCH(1,1)/EGARCH(1,1) specification as follows:

$$\epsilon_t | \Psi_{t-1} = \sigma_t z_t; z_t \sim \mathcal{N}(0,1) \quad (A4a)$$

$$\Rightarrow \epsilon_t^\ell | \Psi_{t-1} = \frac{\sigma_t}{\beta_t^\ell} z_t = \sigma_t^\ell z_t; z_t \sim \mathcal{N}(0,1) \quad (A4b)$$

---

[20] Again, subscript $t$ is day-level indicating day $t$, thus $r_{TT}$ is now $r_t$, and $r_{TT}^\ell$ is $r_t^\ell$.
[21] Technically $|r_t| = \beta_t^\ell |r_t^\ell|$, while for daily returns it is possible that $r_t$ and $r_t^\ell$ may have different signs, for simplicity and without significant economic consequence we assume they have the same sign.

Derived from Equation A4b, we model the error term in the liquidity-adjusted ARMA model of Equation A3a with the standard GARCH(1,1)/EGARCH(1,1) spec as follows:

$$\Rightarrow \epsilon_t^\ell | \Psi_{t-1} = \sigma_t^\ell z_t; z_t \sim \mathcal{N}(0,1) \quad (A5a)$$

$$\text{where:} \quad \sigma_t^\ell = \frac{\sigma_t}{\beta_t^\ell} \quad (A5b)$$

In Equation A5b, the conditional volatility (standard deviation) $\sigma_t^\ell$ of the error term $\epsilon_t^\ell$ is equal to $\sigma_t$ being scaled by $\beta_t^\ell$: when $\beta_t^\ell > 1$, $\sigma_t^\ell < \sigma_t$, and when $\beta_t^\ell < 1$, $\sigma_t^\ell > \sigma_t$. Conceptually, this means that when liquidity is high, the volatility of error term for the liquidity-adjusted return $r_t^\ell$ is lower than that for the regular return $r_t$, while when liquidity is low the opposite is true. The scaling effect of $\beta_t^\ell$ on $\sigma_t^\ell$ is the same as that on $\epsilon_t^\ell$. This gives an additional perspective on that GARCH/EGARCH offers better predictability for the liquidity-adjusted return $r_t^\ell$ than the regular return $r_t$ when liquidity is high.

From Equations 10b and A5b, the liquidity-adjusted GARCH(1,1) specification for the volatility of the error term for liquidity-adjusted return $r_t^\ell$ can be derived from the standard GARCH(1,1) specification as follows:

$$\sigma_t^2 = \omega + a\epsilon_{t-1}^2 + b\sigma_{t-1}^2 \quad (A6a)$$

$$\Rightarrow \beta_t^{\ell^2} \sigma_t^{\ell^2} = \omega + a\beta_{t-1}^{\ell^2} \epsilon_{t-1}^{\ell^2} + b\beta_{t-1}^{\ell^2} \sigma_{t-1}^{\ell^2} \quad (A6b)$$

$$\Rightarrow \sigma_t^{\ell^2} = \frac{\omega}{\beta_t^{\ell^2}} + a\frac{\beta_{t-1}^{\ell^2}}{\beta_t^{\ell^2}} \epsilon_{t-1}^{\ell^2} + b\frac{\beta_{t-1}^{\ell^2}}{\beta_t^{\ell^2}} \sigma_{t-1}^{\ell^2} \quad (A6c)$$

Again, because the extreme values of $\beta_t^\ell$ tend to cluster, when $\beta_t^\ell \gg 1$ or $\beta_t^\ell \ll 1$, $\frac{\beta_{t-1}^\ell}{\beta_t^\ell} \sim 1$, derived from Equation A6c we have the following:

$$\sigma_t^{\ell 2} = \frac{\omega}{\beta_{t-1}^{\ell}{}^2} + a\epsilon_{t-1}^{\ell}{}^2 + b\sigma_{t-1}^{\ell}{}^2 \tag{A7a}$$

$$\Rightarrow \sigma_t^{\ell 2} = \omega + a\epsilon_{t-1}^{\ell}{}^2 + b\sigma_{t-1}^{\ell}{}^2 + \omega\frac{1-\beta_{t-1}^{\ell}{}^2}{\beta_{t-1}^{\ell}{}^2} \tag{A7b}$$

Also, from Equations 10c and A5b, we derive the liquidity-adjusted EGARCH(1,1) specification for the volatility of the error term for liquidity-adjusted return $r_t^\ell$ as:

$$\Rightarrow \log \sigma_t^{\ell 2} = \log \sigma_t^2 - \log \beta_t^{\ell 2} \tag{A8a}$$

$$\Rightarrow = [\omega + bg(Z_{t-1}) + a \log \sigma_{t-1}^2] - \log \beta_t^{\ell 2}$$

$$\Rightarrow = \left[\omega + bg(Z_{t-1}) + a\left(\log \sigma_{t-1}^{\ell}{}^2 + \log \beta_{t-1}^{\ell}{}^2\right)\right] - \log \beta_t^{\ell 2}$$

$$\Rightarrow \log \sigma_t^{\ell 2} = \omega + bg(Z_{t-1}) + a \log \sigma_{t-1}^{\ell}{}^2 + a \log \beta_{t-1}^{\ell}{}^2 - \log \beta_t^{\ell 2} \tag{A8b}$$

Again, taking advantage of that when $\beta_t^\ell \gg 1$ or $\beta_t^\ell \ll 1$, $\frac{\beta_{t-1}^\ell}{\beta_t^\ell} \sim 1$, we rewrite Equation A8b as:

$$\log \sigma_t^{\ell 2} = \omega + bg(Z_{t-1}) + a \log \sigma_{t-1}^{\ell}{}^2 + (a-1) \log \beta_{t-1}^{\ell}{}^2 \tag{A9}$$

Thus, the generalized liquidity-adjusted ARMA-GARCH/EGARCH at extreme liquidity is:

$$ARMA: r_t^\ell = \sum_{i=1}^p \phi_i r_{t-1}^\ell - \sum_{j=1}^q \theta_j \epsilon_{t-j}^\ell + \epsilon_t^\ell \tag{A10a}$$

$$GARCH(1,1): \sigma_t^{\ell 2} = \omega + a\epsilon_{t-1}^{\ell}{}^2 + b\sigma_{t-1}^{\ell}{}^2 + \omega\frac{1-\beta_{t-1}^{\ell}{}^2}{\beta_{t-1}^{\ell}{}^2} \tag{A10b}$$

$$EGARCH(1,1): \log \sigma_t^{\ell 2} = \omega + bg(Z_{t-1}) + a \log \sigma_{t-1}^{\ell}{}^2 + (a-1) \log \beta_{t-1}^{\ell}{}^2 \tag{A10c}$$

where: $g(Z_t) = \theta Z_t + \lambda(|Z_t| - E(|Z_t|))$; $Z_t \sim N(0,1)$

The error and volatility terms of the standard ARMA-GARCH/EGARCH model of Equations A1 are connected to that of the liquidity-adjusted ARMA-GARCH/EGARCH model of Equations A10 as follows:

$$\epsilon_t^\ell = \frac{\epsilon_t}{\beta_t^\ell}; \quad \sigma_t^\ell = \frac{\sigma_t}{\beta_t^\ell} \tag{A11}$$

where: $\epsilon_t|\Psi_{t-1} = \sigma_t z_t; z_t \sim N(0,1)$; $\epsilon_t^\ell|\Psi_{t-1} = \sigma_t^\ell z_t; z_t \sim N(0,1)$

Equations A10 are Equations 11, and Equation A11 is Equations 12 in Subsection 5.

# Table 1 - Descriptive Statistics of Regular Daily Return $r_{TT}$ and Daily Liquidity-adjusted Return $r_{TT}^{\ell}$

This table reports descriptive statistics of daily regular returns ($r_{TT}$) and daily liquidity-adjusted returns ($r_{TT}^{\ell}$), both derived from minute-level returns for each crypto asset over the entire sample period. Panel A shows the descriptive statistics of daily regular returns ($r_{TT}$); Panel B shows the descriptive statistics of daily liquidity-adjusted returns ($r_{TT}^{\ell}$). All ten crypto assets are measured with their trading pairs with Tether or USDT, a "stable coin" pegged to the US dollar, which is regarded as the "risk-free" asset in portfolios with a 0% interest rate in terms of their market values from the Binance API.

| Panel A | daily regular return (return_TT) | | | | | | | | | |
|---|---|---|---|---|---|---|---|---|---|---|
| ticker | ADA | BNB | BTC | ETC | ETH | LINK | LTC | MATIC | XMR | XRP |
| count | 1711 | 1711 | 1711 | 1711 | 1711 | 1711 | 1711 | 1711 | 1711 | 1711 |
| mean | 0.34% | 0.19% | 0.19% | 0.29% | 0.35% | 0.56% | 0.09% | 0.22% | 0.05% | 0.23% |
| std | 9.85% | 6.60% | 3.99% | 7.33% | 6.12% | 7.59% | 5.32% | 9.38% | 6.33% | 8.28% |
| min | -95.05% | -99.11% | -39.50% | -47.54% | -44.60% | -84.85% | -38.53% | -98.91% | -69.85% | -47.22% |
| 25% | -2.62% | -1.71% | -1.42% | -2.32% | -1.88% | -3.04% | -2.34% | -3.71% | -2.42% | -2.08% |
| 50% (median) | 0.04% | 0.13% | 0.03% | 0.02% | 0.11% | 0.31% | 0.06% | -0.06% | 0.26% | 0.05% |
| 75% | 2.53% | 2.17% | 1.75% | 2.40% | 2.36% | 3.35% | 2.41% | 3.52% | 2.58% | 2.09% |
| max | 324.00% | 117.50% | 44.98% | 164.46% | 135.76% | 63.00% | 53.09% | 85.79% | 101.51% | 222.93% |

| Panel B | daily liquidity-adjusted return (return_liquidity_TT) | | | | | | | | | |
|---|---|---|---|---|---|---|---|---|---|---|
| ticker | ADA | BNB | BTC | ETC | ETH | LINK | LTC | MATIC | XMR | XRP |
| count | 1711 | 1711 | 1711 | 1711 | 1711 | 1711 | 1711 | 1711 | 1711 | 1711 |
| mean | 2.05% | 0.62% | 1.09% | 1.05% | 1.48% | 5.49% | 1.44% | 4.46% | 3.53% | 1.46% |
| std | 21.44% | 9.61% | 5.65% | 9.30% | 9.63% | 31.91% | 7.81% | 14.91% | 13.76% | 10.17% |
| min | -97.15% | -99.76% | -40.24% | -74.21% | -54.18% | -62.66% | -52.62% | -98.43% | -61.64% | -65.52% |
| 25% | -1.57% | -1.81% | -1.37% | -2.40% | -1.48% | -1.48% | -1.44% | -2.14% | -1.68% | -1.81% |
| 50% (median) | 0.94% | 0.24% | 0.58% | 0.30% | 0.74% | 1.88% | 0.91% | 1.87% | 1.38% | 0.59% |
| 75% | 4.22% | 2.73% | 2.97% | 3.30% | 3.69% | 6.21% | 3.87% | 7.65% | 4.42% | 3.55% |
| max | 836.28% | 237.64% | 90.23% | 195.92% | 296.94% | 712.51% | 202.21% | 197.13% | 208.84% | 185.92% |

# Table 2 - Descriptive Statistics of Daily Liquidity Premium $\beta_{TT}^{\ell}$

This table reports descriptive statistics of daily liquidity premium Batas ($\beta_{TT}^{\ell}$) for each crypto asset over the entire sample period. The maximum daily liquidity premium Beta is capped at 10. All ten crypto assets are measured with their trading pairs with Tether or USDT, a "stable coin" pegged to the US dollar, which is regarded as the "risk-free" asset in portfolios with a 0% interest rate in terms of their market values from the Binance API.

| | daily liquidity premium Beta (beta_liquidity_TT) | | | | | | | | | |
|---|---|---|---|---|---|---|---|---|---|---|
| ticker | ADA | BNB | BTC | ETC | ETH | LINK | LTC | MATIC | XMR | XRP |
| count | 1711 | 1711 | 1711 | 1711 | 1711 | 1711 | 1711 | 1711 | 1711 | 1711 |
| mean | 1.59 | 1.56 | 1.20 | 1.52 | 1.36 | 1.40 | 1.54 | 1.39 | 1.47 | 1.33 |
| std | 2.16 | 2.09 | 1.70 | 2.08 | 1.89 | 1.97 | 2.06 | 1.91 | 2.14 | 1.91 |
| min | 0.00 | 0.00 | 0.00 | 0.00 | 0.01 | 0.00 | 0.00 | 0.00 | 0.00 | 0.00 |
| 25% | 0.52 | 0.54 | 0.49 | 0.48 | 0.55 | 0.48 | 0.50 | 0.47 | 0.36 | 0.47 |
| 50% (median) | 0.85 | 0.91 | 0.73 | 0.84 | 0.81 | 0.82 | 0.87 | 0.82 | 0.78 | 0.76 |
| 75% | 1.50 | 1.50 | 1.15 | 1.48 | 1.27 | 1.34 | 1.53 | 1.39 | 1.44 | 1.22 |
| max | 10.00 | 10.00 | 10.00 | 10.00 | 10.00 | 10.00 | 10.00 | 10.00 | 10.00 | 10.00 |
| highest days (= max) | 61 | 61 | 36 | 58 | 48 | 50 | 56 | 44 | 58 | 49 |
| % of total days | 3.57% | 3.57% | 2.10% | 3.39% | 2.81% | 2.92% | 3.27% | 2.57% | 3.39% | 2.86% |
| weight in beta | 22.42% | 22.91% | 17.60% | 22.27% | 20.61% | 20.85% | 21.31% | 18.53% | 23.02% | 21.57% |
| highest days (>= mean) | 401 | 410 | 403 | 412 | 382 | 404 | 426 | 428 | 411 | 376 |
| % of total days | 23.44% | 23.96% | 23.55% | 24.08% | 22.33% | 23.61% | 24.90% | 25.01% | 24.02% | 21.98% |
| highest days(>= 1) | 727 | 749 | 525 | 707 | 608 | 653 | 727 | 649 | 635 | 579 |
| % of total days | 42.49% | 43.78% | 30.68% | 41.32% | 35.53% | 38.16% | 42.49% | 37.93% | 37.11% | 33.84% |
| lowest days (<= 0.10) | 62 | 80 | 75 | 70 | 61 | 89 | 59 | 91 | 132 | 60 |
| % of total days | 3.62% | 4.68% | 4.38% | 4.09% | 3.57% | 5.20% | 3.45% | 5.32% | 7.71% | 3.51% |

## Table 3 - Descriptive Statistics of Daily Amount $A_{TT}$

This table reports descriptive statistics of daily trading amounts ($A_{TT}$) for each crypto asset over the entire sample period. All ten crypto assets are measured with their trading pairs with Tether or USDT, a "stable coin" pegged to the US dollar, which is regarded as the "risk-free" asset in portfolios with a 0% interest rate in terms of their market values from the Binance API.

| Panel A | daily amount (amount_TT) | | | | | | | | | |
|---|---|---|---|---|---|---|---|---|---|---|
| ticker | ADA | BNB | BTC | ETC | ETH | LINK | LTC | MATIC | XMR | XRP |
| count | 1711 | 1711 | 1711 | 1711 | 1711 | 1711 | 1711 | 1711 | 1711 | 1711 |
| mean | 77,291,528 | 134,681,195 | 1,034,857,155 | 26,566,835 | 424,890,621 | 38,997,203 | 38,834,250 | 49,073,502 | 5,786,053 | 118,486,429 |
| std | 159,037,469 | 246,273,943 | 1,104,882,901 | 103,018,398 | 479,861,719 | 49,194,715 | 64,743,055 | 123,104,869 | 7,472,868 | 202,862,414 |
| min | 427,225 | 3,856,579 | 33,769,993 | 330,109 | 6,793,237 | 95,884 | 1,197,985 | 316,922 | 165,184 | 1,931,632 |
| 25% | 6,188,484 | 22,143,078 | 265,960,531 | 2,465,658 | 79,497,479 | 7,841,509 | 9,457,161 | 2,818,128 | 1,998,711 | 18,757,429 |
| 50% (median) | 18,634,625 | 47,182,858 | 664,247,922 | 6,685,078 | 278,276,422 | 20,096,882 | 18,238,580 | 17,436,491 | 3,600,098 | 61,074,132 |
| 75% | 69,714,918 | 126,368,869 | 1,406,377,153 | 20,390,447 | 630,101,136 | 53,035,008 | 37,299,821 | 46,731,450 | 6,938,998 | 123,927,705 |
| max | 1,778,526,265 | 2,657,442,021 | 8,633,933,534 | 2,550,632,986 | 5,748,757,771 | 470,743,798 | 723,023,058 | 1,936,750,440 | 104,845,167 | 1,988,415,809 |
| total | 132,245,805,024 | 230,439,525,372 | 1,770,640,592,790 | 45,455,854,264 | 726,987,852,235 | 66,724,213,938 | 66,445,402,371 | 83,964,762,770 | 9,899,936,804 | 202,730,280,464 |
| highest days (25%) | 38 | 45 | 104 | 12 | 102 | 83 | 50 | 25 | 80 | 45 |
| % of total days | 2.22% | 2.63% | 6.08% | 0.70% | 5.96% | 4.85% | 2.92% | 1.46% | 4.68% | 2.63% |
| highest days (50%) | 121 | 147 | 298 | 72 | 305 | 250 | 174 | 115 | 277 | 177 |
| % of total days | 7.07% | 8.59% | 17.42% | 4.21% | 17.83% | 14.61% | 10.17% | 6.72% | 16.19% | 10.34% |
| highest days (75%) | 307 | 388 | 629 | 277 | 605 | 539 | 506 | 354 | 650 | 493 |
| % of total days | 17.94% | 22.68% | 36.76% | 16.19% | 35.36% | 31.50% | 29.57% | 20.69% | 37.99% | 28.81% |
| Panel B | daily amount (amount_TT) as percentage of raw daily amount (without treatments of wash trades) | | | | | | | | | |
| ticker | ADA | BNB | BTC | ETC | ETH | LINK | LTC | MATIC | XMR | XRP |
| mean | 40.33% | 43.00% | 46.54% | 37.40% | 43.72% | 40.08% | 39.75% | 39.55% | 34.43% | 43.75% |
| min | 29.56% | 30.95% | 32.93% | 27.37% | 32.24% | 26.47% | 28.78% | 28.80% | 27.15% | 28.74% |
| median | 40.35% | 42.85% | 46.37% | 37.33% | 43.68% | 40.42% | 39.49% | 39.88% | 33.97% | 44.21% |
| max | 54.02% | 53.60% | 57.62% | 50.91% | 53.15% | 51.73% | 51.53% | 53.58% | 47.88% | 54.37% |

**Table 4 - Descriptive Statistics of Daily Regular Minute-Level (intraday) Variance ($\sigma^2_{TT}$) and Daily Liquidity-Adjusted Minute-Level (intraday) Variances ($\sigma^{2\ell}_{TT}$)**

This table reports descriptive statistics of daily regular minute-level (intraday) return variances ($\sigma^2_{TT}$) and daily liquidity-adjusted minute-level (intraday) return variances ($\sigma^{2\ell}_{TT}$) for each crypto asset over the entire sample period. Panel A shows the descriptive statistics of daily regular minute-level (intraday) variances ($\sigma^2_{TT}$); Panel B shows the descriptive statistics of daily liquidity-adjusted minute-level (intraday) variances ($\sigma^{2\ell}_{TT}$). All ten crypto assets are measured with their trading pairs with Tether or USDT, a "stable coin" pegged to the US dollar, which is regarded as the "risk-free" asset in portfolios with a 0% interest rate in terms of their market values from the Binance API.

| Panel A | daily regular minute-level (intraday) variance ($\sigma^2_{TT}$) | | | | | | | | | |
|---|---|---|---|---|---|---|---|---|---|---|
| ticker | ADA | BNB | BTC | ETC | ETH | LINK | LTC | MATIC | XMR | XRP |
| count | 1711 | 1711 | 1711 | 1711 | 1711 | 1711 | 1711 | 1711 | 1711 | 1711 |
| mean | 0.36% | 0.26% | 0.15% | 0.40% | 0.22% | 0.51% | 0.31% | 0.79% | 0.30% | 0.37% |
| std | 0.76% | 0.63% | 0.36% | 0.84% | 0.53% | 0.98% | 0.56% | 1.81% | 0.51% | 0.94% |
| min | 0.02% | 0.01% | 0.00% | 0.02% | 0.00% | 0.01% | 0.01% | 0.01% | 0.04% | 0.01% |
| 25% | 0.12% | 0.07% | 0.04% | 0.11% | 0.06% | 0.15% | 0.11% | 0.17% | 0.11% | 0.08% |
| 50% (median) | 0.20% | 0.13% | 0.08% | 0.20% | 0.12% | 0.28% | 0.18% | 0.38% | 0.19% | 0.14% |
| 75% | 0.35% | 0.23% | 0.15% | 0.38% | 0.23% | 0.51% | 0.32% | 0.78% | 0.34% | 0.29% |
| max | 22.71% | 13.11% | 9.48% | 18.12% | 11.06% | 21.04% | 11.09% | 45.38% | 11.14% | 19.09% |

| Panel B | daily liquidity-adjusted minute-level (intraday) variance ($\sigma^{2\ell}_{TT}$) | | | | | | | | | |
|---|---|---|---|---|---|---|---|---|---|---|
| ticker | ADA | BNB | BTC | ETC | ETH | LINK | LTC | MATIC | XMR | XRP |
| count | 1711 | 1711 | 1711 | 1711 | 1711 | 1711 | 1711 | 1711 | 1711 | 1711 |
| mean | 0.73% | 0.74% | 0.37% | 0.74% | 0.53% | 0.94% | 0.60% | 1.48% | 0.50% | 0.87% |
| std | 3.80% | 7.42% | 1.31% | 3.29% | 2.16% | 3.13% | 1.95% | 6.97% | 1.73% | 3.13% |
| min | 0.02% | 0.02% | 0.00% | 0.03% | 0.00% | 0.02% | 0.01% | 0.01% | 0.06% | 0.01% |
| 25% | 0.16% | 0.13% | 0.10% | 0.13% | 0.13% | 0.23% | 0.14% | 0.30% | 0.13% | 0.15% |
| 50% (median) | 0.29% | 0.24% | 0.18% | 0.26% | 0.24% | 0.43% | 0.27% | 0.57% | 0.23% | 0.27% |
| 75% | 0.58% | 0.46% | 0.33% | 0.56% | 0.46% | 0.84% | 0.53% | 1.21% | 0.46% | 0.60% |
| max | 145.97% | 294.18% | 34.42% | 115.23% | 59.20% | 89.46% | 48.66% | 255.73% | 61.00% | 72.40% |

**Table 5 - Adam-Fuller Tests of Stationary on Daily Regular Return ($r_{TT}$) and Daily Liquidity-Adjusted Return ($r_{TT}^{\ell}$) with ARMA-GARCH Forecast Accuracy (RMSE)**

This table reports the Adam-Fuller stationary test (ADF) results on daily regular returns ($r_{TT}$) and daily liquidity-adjusted returns ($r_{TT}^{\ell}$). Panel A reports the full-sample test results and out-of-sample ARMA-GARCH forecast accuracy (RMSE) for daily regular returns ($r_{TT}$); Panel B reports the full-sample test results and out-of-sample ARMA-GARCH forecast accuracy (RMSE) for daily liquidity-adjusted returns ($r_{TT}^{\ell}$). All ten crypto assets are measured with their trading pairs with Tether or USDT, a "stable coin" pegged to the US dollar, which is regarded as the "risk-free" asset in portfolios with a 0% interest rate in terms of their market values from the Binance API.

| Panel A | daily regular return ($r_{TT}$) | | | | | | | | | |
|---|---|---|---|---|---|---|---|---|---|---|
| | ADA_USDT | BNB_USDT | BTC_USDT | ETC_USDT | ETH_USDT | LINK_USDT | LTC_USDT | MATIC_USDT | XMR_USDT | XRP_USDT |
| ADF critical values | 1%: -4.379; 5%: -3.836; 10%: -3.556 | | | | | | | | | |
| ADF test | -41.734 | -26.930 | -42.865 | -6.778 | -18.805 | -41.763 | -43.359 | -9.215 | -46.930 | -40.259 |
| p_value | 0.00 | 0.00 | 0.00 | 0.00 | 0.00 | 0.00 | 0.00 | 0.00 | 0.00 | 0.00 |
| used_lag | 0 | 1 | 0 | 24 | 4 | 0 | 0 | 15 | 0 | 0 |
| nobs | 1710 | 1709 | 1710 | 1686 | 1706 | 1710 | 1710 | 1695 | 1710 | 1710 |
| ic_best | -3107 | -4636 | -6081 | -4090 | -4840 | -4356 | -5105 | -3537 | -4712 | -4540 |
| RMSE (1,346 days out-of-sample) | 0.1374 | 0.1116 | 0.0535 | 0.1272 | 0.0748 | 0.0610 | 0.0570 | 0.0922 | 0.0615 | 0.1253 |

| Panel B | daily liquidity-adjusted return ($r_{TT}^{\ell}$) | | | | | | | | | |
|---|---|---|---|---|---|---|---|---|---|---|
| | ADA_USDT | BNB_USDT | BTC_USDT | ETC_USDT | ETH_USDT | LINK_USDT | LTC_USDT | MATIC_USDT | XMR_USDT | XRP_USDT |
| ADF critical values | 1%: -4.379; 5%: -3.836; 10%: -3.556 | | | | | | | | | |
| ADF test | -41.074 | -27.107 | -28.442 | -5.712 | -19.342 | -11.027 | -9.160 | -6.671 | -2.142 | -11.740 |
| p_value | 0.00 | 0.00 | 0.00 | 0.00 | 0.00 | 0.00 | 0.00 | 0.00 | 0.76 | 0.00 |
| used_lag | 0 | 1 | 1 | 24 | 4 | 24 | 12 | 21 | 25 | 8 |
| nobs | 1710 | 1709 | 1709 | 1686 | 1706 | 1686 | 1698 | 1689 | 1685 | 1702 |
| ic_best | -403 | -3221 | -4914 | -3386 | -3190 | -3571 | -3824 | -2391 | -3173 | -3289 |
| RMSE (1,346 days out-of-sample) | 0.4616 | 0.1039 | 0.1215 | 0.1443 | 0.1325 | 0.2092 | 0.1112 | 0.1178 | 0.0973 | 0.1838 |

## Table 6 – Performance Comparisons

This table compares the performance of eight portfolios, including four benchmark portfolios (1 – 4), two TMV portfolios (5, 7) and two LAMV portfolios (6, 8). The risk-free rate ($R_f$) is assumed to be zero. The abbreviations of "equ", "mkt", "blq", "blq_inv", "MV_rr", "MV_rrlq", "MV_arga_rr" and "MV_arga_rrlq" stand for "equal weight", "market weight", "liquidity weight", "inverse liquidity weight", "MV without liquidity adjustment", "MV with liquidity adjustment", "MV without liquidity adjustment, ARMA-GARCH/EGARCH-enhanced" and "MV with liquidity Adjustment, ARMA-GARCH/EGARCH-enhanced," respectively.

| Panel A | Portfolios with Wash Trades Removed | | | |
|---|---|---|---|---|
| Portfolios | benchmark | | | |
| Portfolio Number | 1 | 2 | 3 | 4 |
| Portfolio Description | equ | mkt | blq | blq_inv |
| Annualized Return | 138.26% | 110.39% | 161.44% | 86.62% |
| Annualized Standard Deviation | 91.12% | 86.83% | 112.52% | 89.30% |
| Annualized Sharpe Ratio ($R_f$=0%) | **1.52** | **1.27** | **1.43** | **0.97** |
| Portfolios | regular daily return | | liquidity-adjusted daily return | |
| Portfolio Number | 5 | | 6 | |
| Portfolio Description | MV_rr | | MV_rrlq | |
| Annualized Return | 153.26% | | 99.48% | |
| Annualized Standard Deviation | 107.02% | | 97.61% | |
| Annualized Sharpe Ratio ($R_f$=0%) | **1.43** | | **1.02** | |
| Portfolio Number | 7 | | 8 | |
| Portfolio Description | MV_arga_rr | | MV_arga_rrlq | |
| Annualized Return | 74.04% | | 144.54% | |
| Annualized Standard Deviation | 91.55% | | 95.59% | |
| Annualized Sharpe Ratio ($R_f$=0%) | **0.81** | | **1.51** | |

| Panel B | Portfolios with Wash Trades Retained | | | |
|---|---|---|---|---|
| Portfolios | benchmark | | | |
| Portfolio Number | 1 | 2 | 3 | 4 |
| Portfolio Description | equ | mkt | blq | blq_inv |
| Annualized Return | 138.59% | 109.75% | 111.84% | 105.21% |
| Annualized Standard Deviation | 91.08% | 87.24% | 93.23% | 93.14% |
| Annualized Sharpe Ratio ($R_f$=0%) | **1.52** | **1.26** | **1.20** | **1.13** |
| Portfolios | regular daily return | | liquidity-adjusted daily return | |
| Portfolio Number | 5 | | 6 | |
| Portfolio Description | MV_rr | | MV_rrlq | |
| Annualized Return | 149.77% | | 149.11% | |
| Annualized Standard Deviation | 107.76% | | 96.81% | |
| Annualized Sharpe Ratio ($R_f$=0%) | **1.39** | | **1.54** | |
| Portfolio Number | 7 | | 8 | |
| Portfolio Description | MV_arga_rr | | MV_arga_rrlq | |
| Annualized Return | 78.86% | | 198.66% | |
| Annualized Standard Deviation | 92.49% | | 96.48% | |
| Annualized Sharpe Ratio ($R_f$=0%) | **0.85** | | **2.06** | |

| Panel C | Portfolios | | | |
|---|---|---|---|---|
| Portfolios | benchmark | | | |
| Portfolio Number | 1 | 2 | 3 | 4 |
| Portfolio Description | equ | mkt | blq | blq_inv |
| a: wash trades removed - annualized Sharpe Ratio (Rf=0%) | 1.52 | 1.27 | **1.43** | **0.97** |
| b: wash trades retained - annualized Sharpe Ratio (Rf=0%) | 1.52 | 1.26 | **1.20** | **1.13** |
| Portfolios | regular daily return | | liquidity-adjusted daily return | |
| Portfolio Number | 5 | | 6 | |
| Portfolio Description | MV_rr | | MV_rrlq | |
| a: wash trades removed - annualized Sharpe Ratio (Rf=0%) | 1.43 | | **1.02** | |
| b: wash trades retained - annualized Sharpe Ratio (Rf=0%) | 1.39 | | **1.54** | |
| Portfolio Number | 7 | | 8 | |
| Portfolio Description | MV_arga_rr | | MV_arga_rrlq | |
| a: wash trades removed - annualized Sharpe Ratio (Rf=0%) | 0.81 | | **1.51** | |
| b: wash trades retained - annualized Sharpe Ratio (Rf=0%) | 0.85 | | **2.06** | |

# Figure 1 – Histograms of Daily Liquidity Premium $\beta^{\ell}_{TT}$

This figure provides the distributions (histograms) of daily liquidity premium Betas ($\beta^{\ell}_{TT}$) for each crypto asset over the entire sample period. The maximum value of $\beta^{\ell}_{TT}$ is capped at 10.

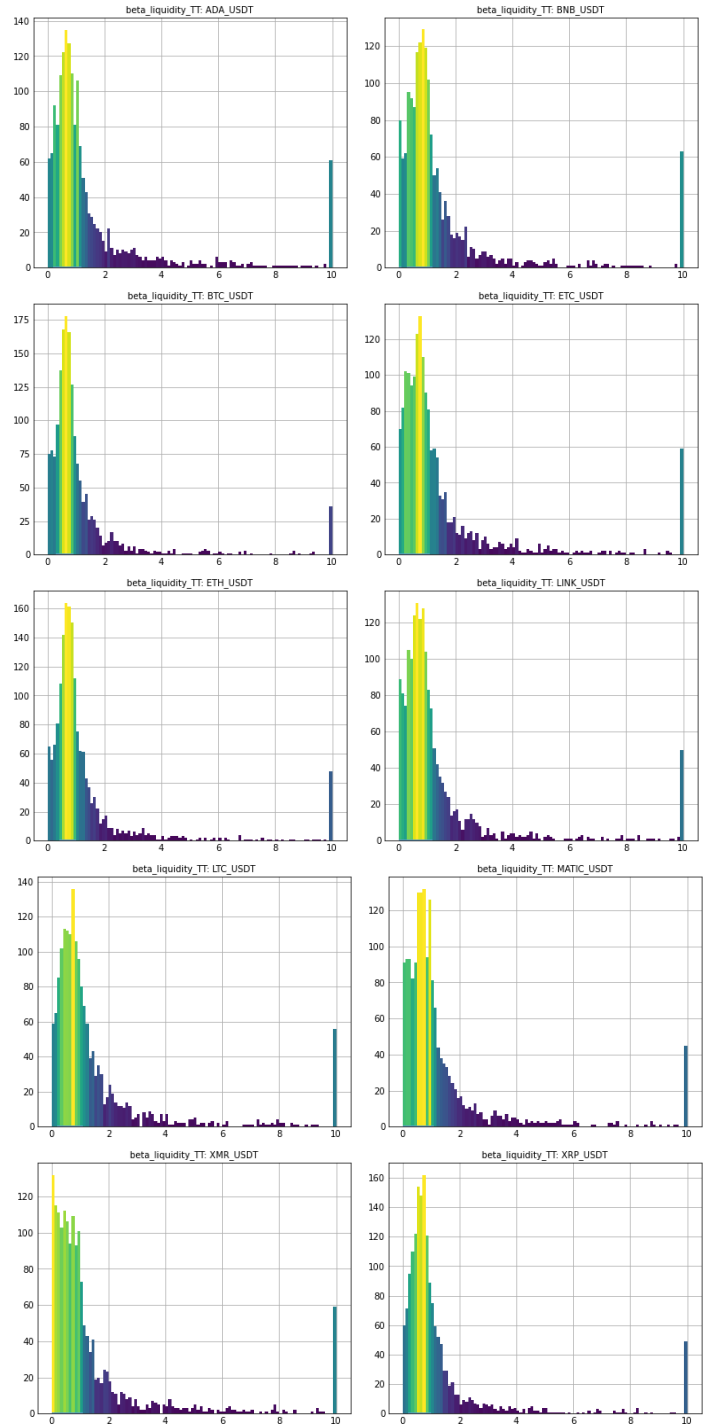

# Figure 2 – Histograms of Daily Amount $A_{TT}$

This figure provides the distributions (histograms) of the daily amounts ($A_{TT}$) for each crypto asset over the entire sample period.

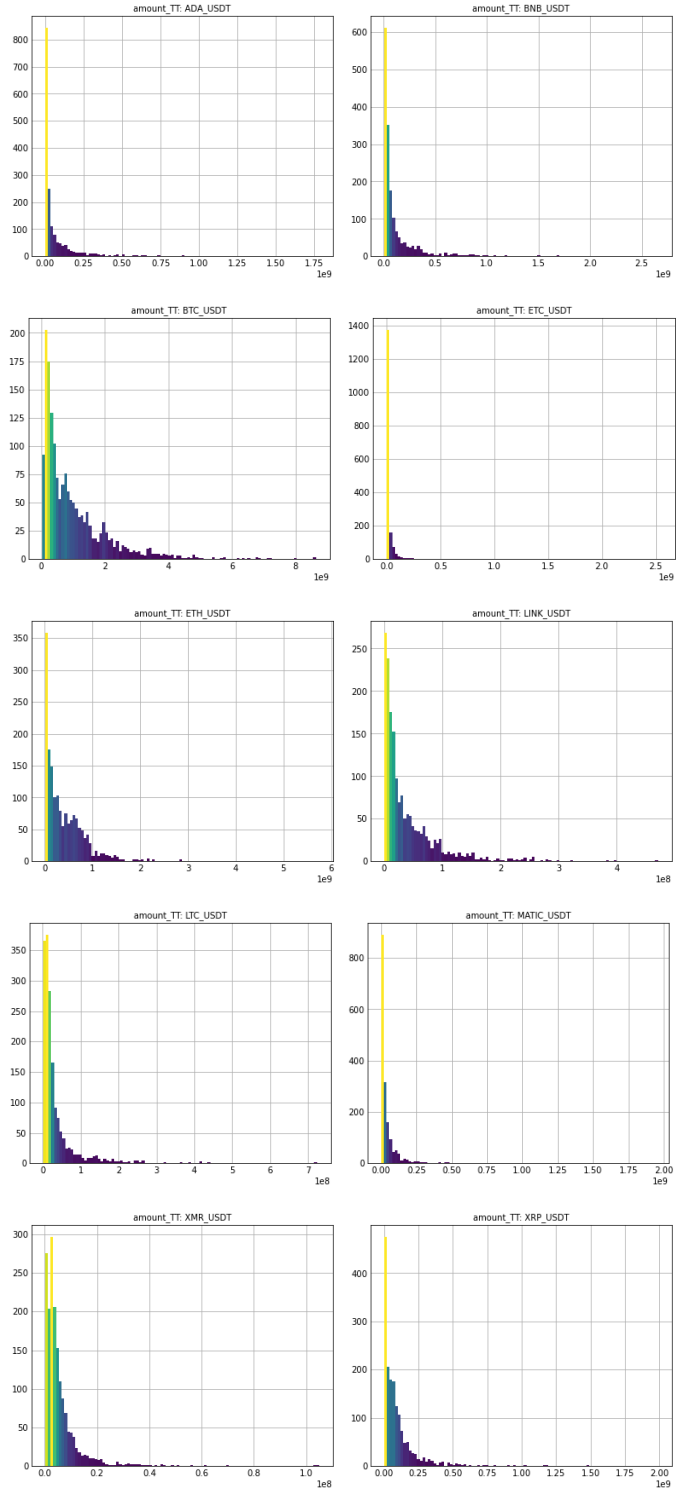

**Figure 3 – Scatter Plots of $r_{TT}$- $\sigma_{TT}$ and $r_{TT}^{\ell}$-$\sigma_{TT}^{\ell}$: Overlay Comparisons**

This figure plots daily regular returns and minute-level (intraday) standard deviations ($r_{TT}$- $\sigma_{TT}$), and daily liquidity-adjusted returns and minute-level (intraday) standard deviations ($r_{TT}^{\ell}$-$\sigma_{TT}^{\ell}$) for each crypto asset over the entire sample period for comparisons.

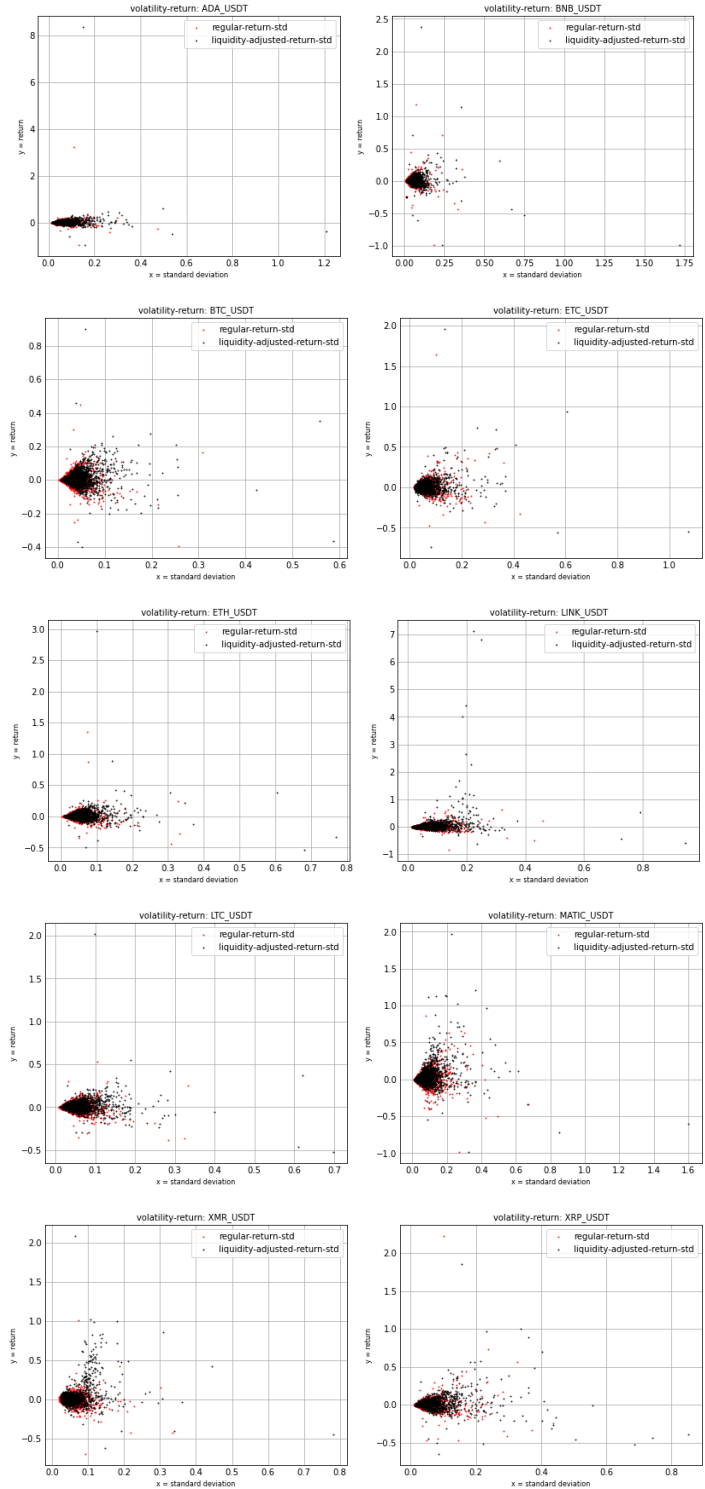

# Figure 4 – Three-Dimensional Distribution $r_{TT}$- $\sigma_{TT}$-$\beta^{\ell}_{TT}$

This figure provides 3-dimensional distributions of the daily regular return ($r_{TT}$), daily regular minute-level (intraday) standard deviation ($\sigma_{TT}$) and daily liquidity premium Beta ($\beta^{\ell}_{TT}$) for each crypto asset over the entire sample period.

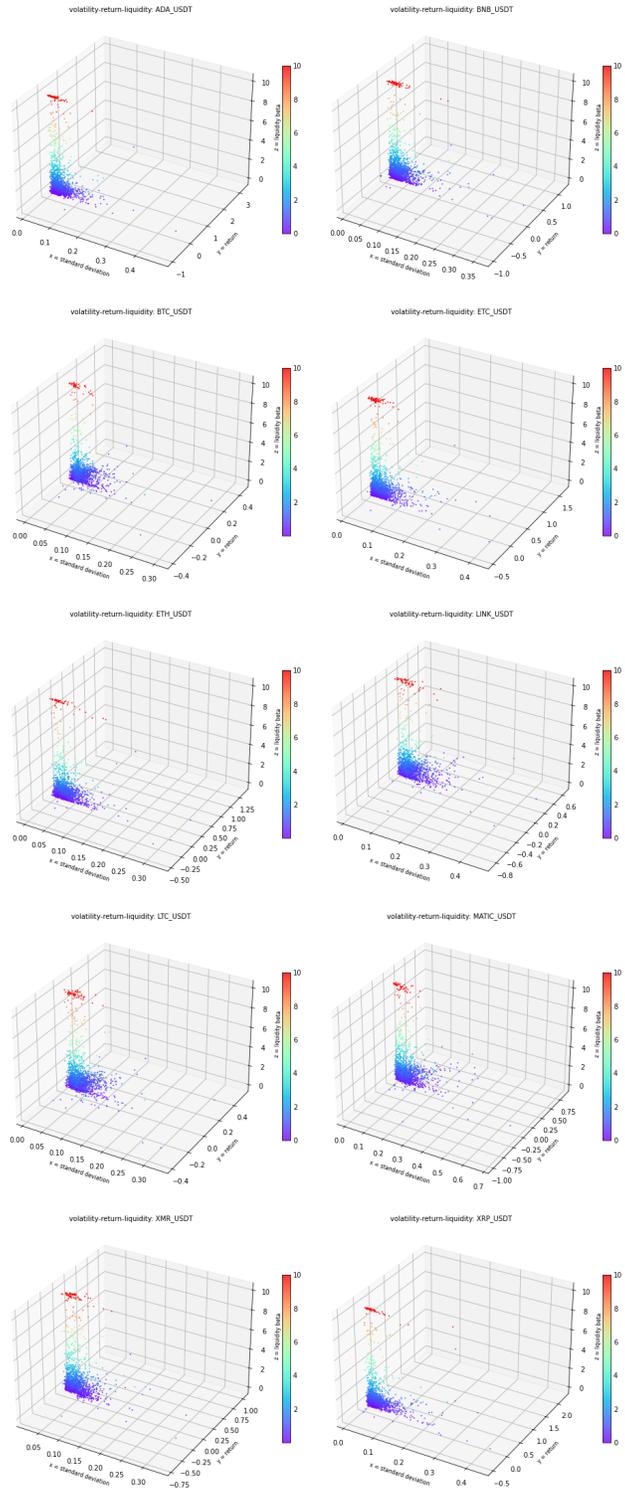

# Figure 5 – Time Series of Daily Amount $A_{TT}$

This figure plots time series of the daily trading amount ($A_{TT}$) for each crypto asset over the entire sample period.

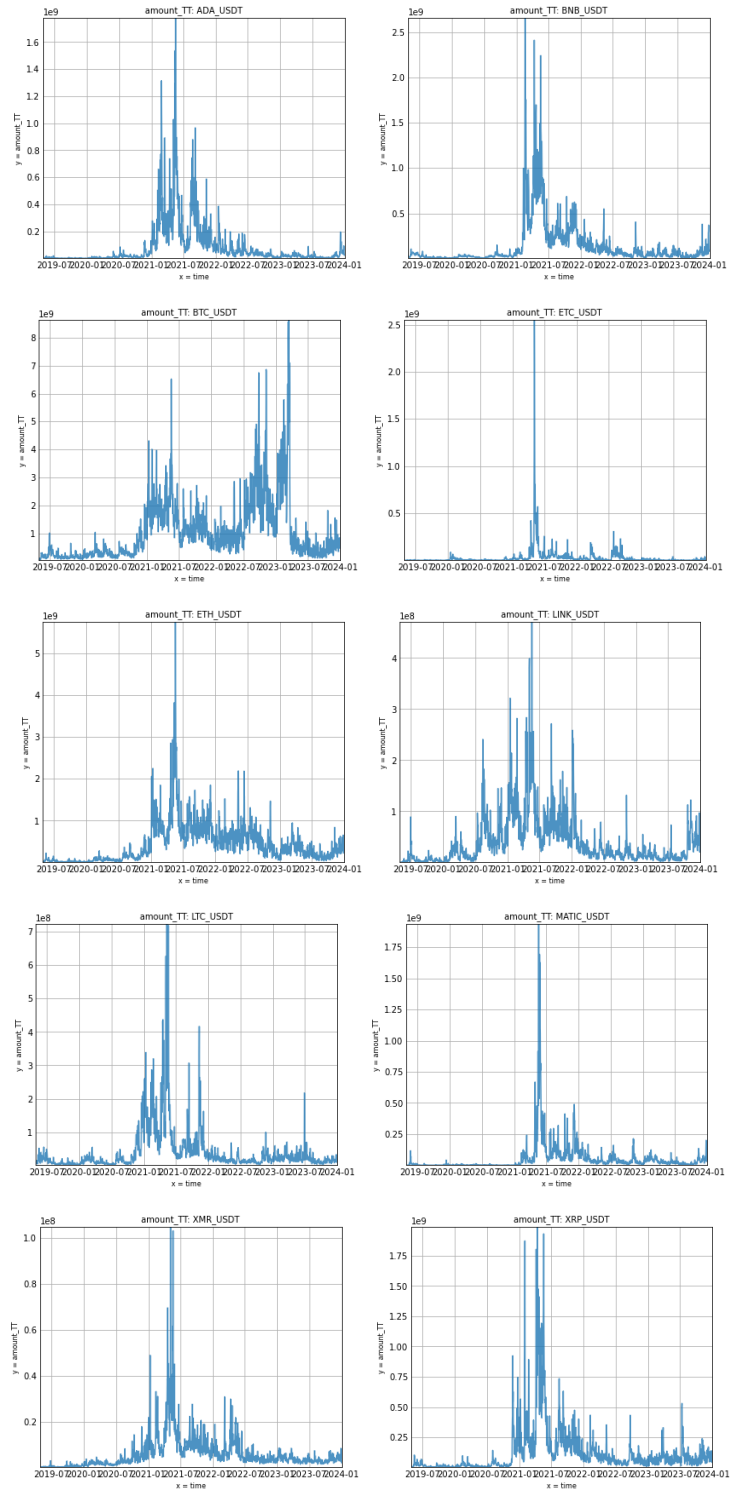

**Figure 6 – Performance of Portfolios with Wash Trades Removed**

This figure provides visual comparisons of portfolio performances. Plot A presents the equity curves of the four benchmark portfolios. Panel B presents the equity curves of the four MV portfolios.

Plot A – equity curves of four benchmark portfolios

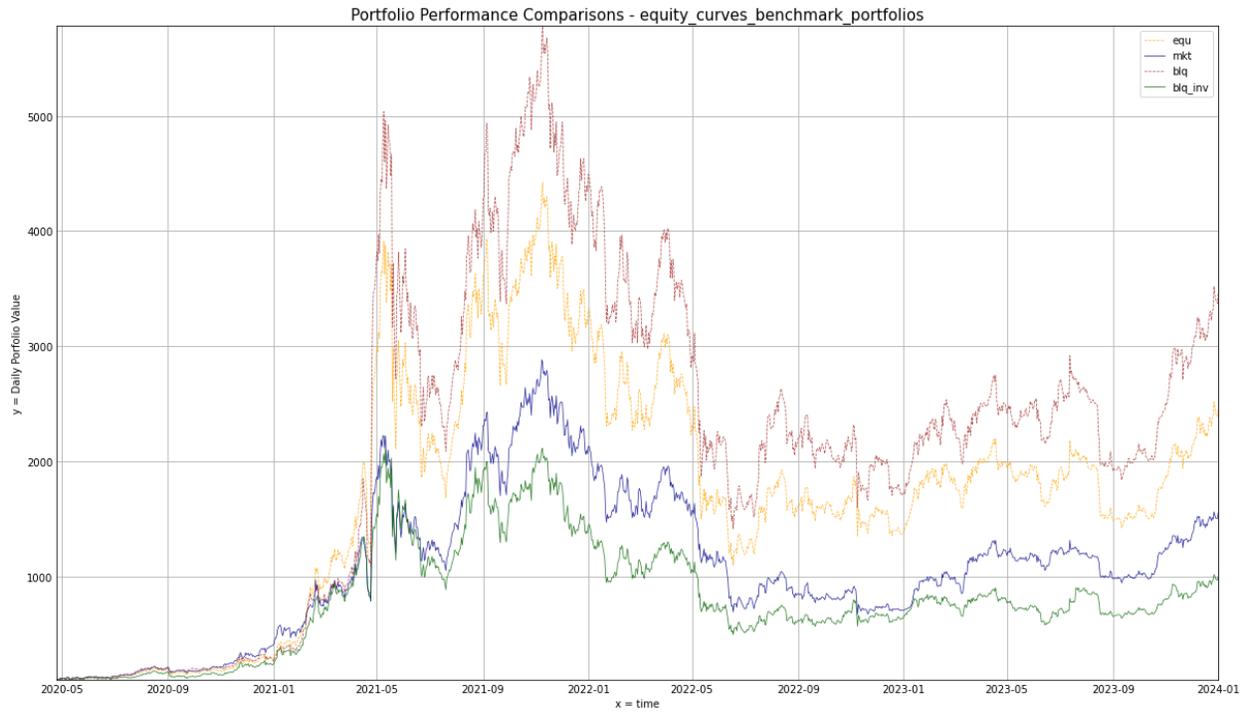

Plot B – equity curves of four MV portfolios

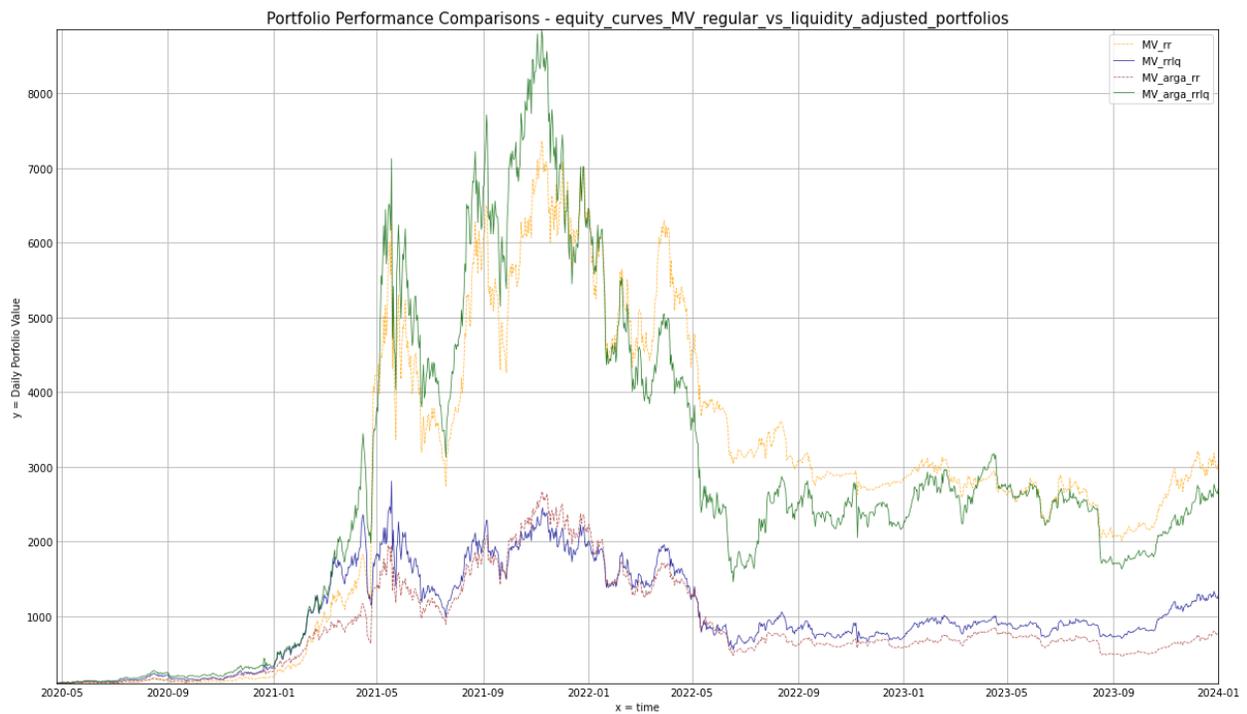

**Figure 7 – Performance of Portfolios with Wash Trades Retained**

This figure provides visual comparisons of portfolio performances. Plot A presents the equity curves of the four benchmark portfolios. Panel B presents the equity curves of the four MV portfolios.

Plot A – equity curves of four benchmark portfolios

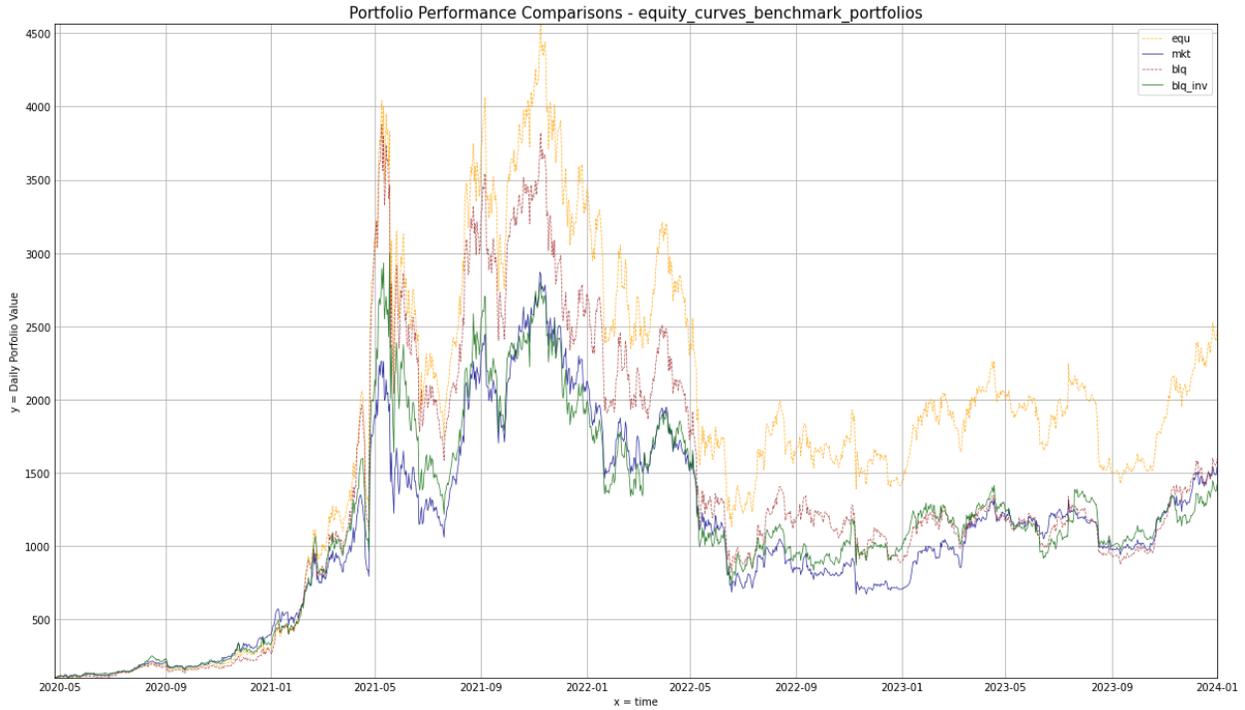

Plot B – equity curves of four MV portfolios.

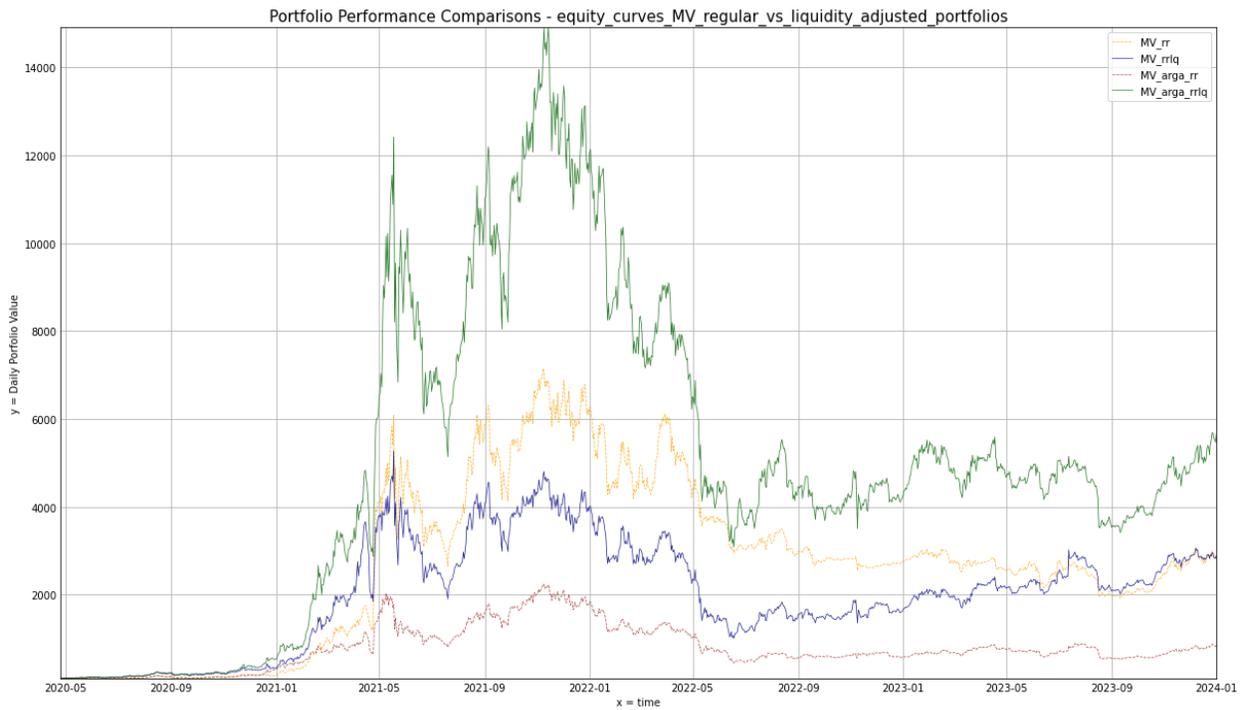